\documentclass[mathpazo]{cicp}
\RequirePackage[dvips]{graphicx}

\def\reals{{\mathbb{R}}}
\renewcommand\vec[1]{\ensuremath\boldsymbol{#1}}
\usepackage[usenames,dvips]{color}

\begin{document}
\title{Self-gravitational force calculation of infinitesimally thin gaseous disks}

\author[Yen CC et.~al.]{C.C. Yen\affil{1,2}\comma\corrauth,
R.E. Taam\affil{2,5}, Ken H.C. Yeh\affil{2,4}, and  K.C. Jea\affil{1,3}}
\address{\affilnum{1}\ Department of Mathematics, Fu Jen Catholic University, 
New Taipei City, Taiwan. \\
\affilnum{2}\ Institute of Astronomy \& Astrophysics, Academia Sinica, 
Taipei, Taiwan.\\
\affilnum{3}\ Graduate Institute of Applied Science and Engineering, Fu Jen Catholic University, 
New Taipei City, Taiwan.\\
\affilnum{4}\
Department of Physics and Astronomy, University of British Columbia, Vancouver, Canada.\\
\affilnum{5}\
Department of Physics and Astronomy, Northwestern University,       
       2145 Sheridan Road, Evanston, IL 60208}
\emails{{\tt yen@math.fju.edu.tw} (C.C. Yen)}


\begin{abstract}
A thin gaseous disk has often been investigated in the context of various phenomena 
in galaxies, which point to the existence of starburst rings and dense circumnuclear 
molecular disks.  The effect of self-gravity of the gas in the $2D$ disk can be important in 
confronting observations and numerical simulations in detail.
For use in such applications, a new method for the calculation of the gravitational 
force of a $2D$  disk is presented.  Instead of solving the complete potential 
function problem, we calculate the force in infinite planes in Cartesian and polar 
coordinates by a reproducing kernel method.
Under the limitation of a $2D$ disk,
we specifically represent the force as a double summation 
of a convolution of the surface density
and a fundamental kernel and employ a fast Fourier transform technique.
In this method, the entire computational complexity can be reduced from $O(N^2\times N^2)$
to $O(N^2(\log_2 N)^2)$, where $N$ is the number of zones in one dimension.
This approach does not require softening.  The proposed method is similar to 
a spectral method, but without the necessity of imposing 
a periodic boundary condition.  We further show this approach is of near second 
order accuracy for a smooth surface density in a Cartesian coordinate system.
\end{abstract}


\maketitle

\section{Introduction}
The potential $\Phi$ of a given distribution of density $\rho$ in $\reals^3$ satisfies the Poisson equation,
\begin{eqnarray}
\label{DPG}
\Delta \Phi(\vec{x}) = 4\pi G \rho(\vec{x})=f(\vec{x}),\quad \vec{x}\in \reals^3,
\end{eqnarray}
where $G$ is the gravitational constant 
and $\vec{x}=(x,y,z)$ is the position.
Without loss of generality, we may assume that the gravitational constant $G=1$.
Provided that the density profile has a continuous second derivative 
with respect to the spatial coordinates, the potential is smooth. 
In this situation, the numerical approach for 
solving the potential via (\ref{DPG}) 
by the finite difference method is adopted.
Artificial boundary conditions are imposed in the numerical approach for solving 
(\ref{DPG}) because the boundary condition is 
\begin{eqnarray}
\label{DPGBC}
\lim_{|\vec{x}|\to \infty}\Phi(\vec{x})=0.
\end{eqnarray}  
The Poisson equation is intrinsically 3-dimensional, and the calculation of the 
potential can be computationally prohibitive.
A possible solution to reduce the computation time is to apply
the multigrid method~\cite{JZhang,AJRoberts01},
but the computational complexity is $O(N^3)$, 
where $N$ is the number of zones in one dimension. 

The solution of (\ref{DPG}) can be represented in terms of the fundamental solution, 
$\displaystyle \frac{1}{4\pi}{\cal K}(\vec{x})$, where
\begin{eqnarray*}
{\cal K}(\vec{x})=\frac{1}{\sqrt{x^2+y^2+z^2}},
\end{eqnarray*} 
as 
\begin{eqnarray}
\label{IPG}
\Phi(x,y,z)=-\int\!\!\!\int\!\!\!\int 
{\cal K}(\bar x-x,\bar y - y, \bar z-z)f(\bar x,\bar y,\bar z)d\bar x d\bar y d\bar z.
\end{eqnarray}
The above formula is preferable to (\ref{DPG}) when the density is not smooth. 
The potential can be solved via the integral equation in ~(\ref{IPG}).
Spectral methods are a common method of choice and a review article has recently been 
written by Shen and Wang \cite{JieShen09}, 
describing work on the analysis and application of these methods in unbounded domains. 
The difficulties encountered in the numerical approach for solving (\ref{DPG}) or 
(\ref{IPG}) are related to the extent of the domain $\reals^3$ and the density 
which can be singular.   

In this paper, we consider the density represented by 
\begin{eqnarray}
\label{SigmaDef}
\rho(\vec{x})=\sigma(x,y)\delta(z),
\end{eqnarray}
where $\sigma(x,y)$ is so-called surface density equal to 
\begin{eqnarray}
\label{SigmaInt}
\sigma(x,y)=\int \rho(\vec{x})dz.
\end{eqnarray}
We restrict our attention to calculating the forces directly
for the surface density of compact supports. 

For an infinitesimally thin gaseous disk, the multigrid method, which is 
intrinsically suited for 3D problems, cannot be reduced for the two dimensional problem 
we consider in this paper.  
The spectral method using Fourier basis functions
on a two dimensional space artificially imposes
the assumption of periodic boundary conditions.
This is not realistic for the long range gravitational force calculations.
A direct method without the periodic assumption requires
a softening parameter technqiue, but the accuracy is reduced simultaneously.
A method is proposed which is of linear complexity, without artificial boundary conditions,
and near second order accuracy.

This paper is organized as follows.
The framework and assumption are presented in Section \ref{secFramework}. 
Sections \ref{secCartesian} and \ref{secPolar} describe the numerical methods for Cartesian and polar coordinates,
respectively. Section \ref{secAccurcy} demonstrates the order of accuracy of the proposed methods 
as verified by a family of finite disks (e.g., $D_2$ disk; \cite{ESchulz}) and a disk of a pair of spirals.
A comparison with several existing methods is also presented in that section.
Finally, the discussion and conclusion are given in section \ref{secConlusion}.

\section{Framework and assumption}
\label{secFramework}
The evolution of a thin disk is of fundamental  interest in astrophysics and the effect of the 
self-gravity of gas therein may be important in modeling observed phenomena in detail.  
This paper presents a numerical method for 
solving the self-gravitating forces in Cartesian and polar coordinates, which can be 
used in modeling infinitesimally thin disks in galaxies and protostellar systems 
\cite{HuiZhang08}.

The self-gravitating force can be determined 
by taking derivatives of the potential function which satisfies the Poisson equation in (\ref{DPG}).
However, the calculation of the potential (\ref{DPG}) is on an unbounded domain and
the solution in a finite region requires the imposition of artificial boundary conditions.
The solution of Poisson's equation with variable coefficients and
Dirichlet boundary conditions on a two dimensional irregular domain is one of second 
order\cite{HansJohansen98}.

Let us confine our attention to the density in an infinitesimally thin disk as
defined in (\ref{SigmaDef}) and (\ref{SigmaInt}).
Here, we focus on the self-gravitating force computation.  
The approach presented in this paper is to directly calculate the self-gravitating force
by expressing the potential function as a type of a convolution of the surface density 
and the fundamental kernel and taking the derivative of the potential function. 
This approach is similar to the spectral method, but less restrictive. 
Trigonometric bases functions and the artificial periodic boundary conditions 
are used for the spectral method, but are not required in the proposed approach here.  
 
A uniform grid discretization in Cartesian coordinates
and a linear approximation of the surface density on each cell are used 
to reduce the computational time and increase the accuracy of 
the numerical solution, respectively.
Similarly, for polar coordinates, a logarithmic grid discretization
is used instead of a uniform grid discretization.
Based on the discretization and approximation, 
the self-gravitating force is written as a convolution form of double summations.
It is known that the calculation of  convolution form can be accelerated
by the use of a fast Fourier transform (FFT), see Appendix B.
Employing the FFT, the computational complexity is reduced from $O(N^4)$
to $O((N\log_2 N)^2)$, 
where $N$ is the number of zones in one direction. 
The linear approximation also leads to 
an order of convergence that is near second order $O(h^2)$,  
where the size of a zone $h = O(1/N)$.

\section{Self-gravitating force calculation in Cartesian coordinates}
\label{secCartesian}

In this section, we describe the method in detail. 
The potential function $\Phi$ of (\ref{DPG}) can be expressed as
\begin{eqnarray*}
\Phi(x,y,z)=-\int\!\!\!\int\!\!\!\int 
{\cal K}(\bar x-x,\bar y - y, \bar z-z)\rho(\bar x,\bar y,\bar z)d\bar x d\bar y d\bar z,
\end{eqnarray*}
where
$\displaystyle {\cal K}(x,y,z)=\frac{1}{\sqrt{x^2+y^2+z^2}}.$
By (\ref{SigmaDef}), 
the forces on the disk in the $x$-direction and the $y$-direction become
\begin{eqnarray}
\label{xforce}
\frac{\partial}{\partial x} \Phi(x,y,0)
=\int\!\!\!\int \frac{\partial}{\partial x}{\cal K}(\bar x-x,\bar y-y,0)\sigma(\bar x,\bar y)
d\bar x d\bar y
\end{eqnarray}
and 
\begin{eqnarray}
\label{yforce}
\frac{\partial}{\partial y} \Phi(x,y,0)
=\int\!\!\!\int \frac{\partial}{\partial y}{\cal K}(\bar x-x,\bar y-y,0)
\sigma(\bar x,\bar y) d\bar x d\bar y.
\end{eqnarray}
We calculate (\ref{xforce}) and (\ref{yforce}) by a numerical approach.
Here, we focus on the derivation of the force calculation in the $x$-direction.
The force in the $y$-direction is obtained in a similar manner (see Appendix A).

Since the support of the surface density is compact, 
contained in a domain $D=[-M,M]\times [-M,M]$ for some number $M>0$,
we discretize the region uniformly as follows. 
Given a positive integer $N$, we define
$\Delta x=2M/N$, $\Delta y =\Delta x$, 
$x_{i+1/2} = -M+i\Delta x$, $y_{j+1/2} = -M + j \Delta y$,
where $i,j=0,\ldots,N$. We further define the center of the cell $D_{ij}=[x_{i-1/2},x_{i+1/2}]
\times [y_{j-1/2},y_{j+1/2}]$ as  
$x_i= (x_{i-1/2}+x_{i+1/2})/2$ and 
$y_j= (y_{j-1/2}+y_{j+1/2})/2$,
where $i,j=1,\ldots,N$. 
Hence, the domain $D$ is discretized into the $N^2$ cells.

The forces in the $x$-direction and the $y$-direction at the center of cells are 
\begin{eqnarray}
\label{eqnFxFy}
F^x_{i,j}=\frac{\partial}{\partial x}\Phi(x_i,y_j,0),\mbox{\quad and \quad}
F^y_{i,j}=\frac{\partial}{\partial y}\Phi(x_i,y_j,0).
\end{eqnarray}
The surface density 
$\sigma$ on $D_{i,j}$ in (\ref{xforce}) is linearly approximated by 
\begin{eqnarray}
\label{sigmaapro}
\sigma(\bar x,\bar y)\approx \sigma_{i,j} 
+\delta^x_{i,j}(\bar x-x_i) + \delta^y_{i,j}(\bar y-y_j),
\end{eqnarray}
where $\sigma_{i,j}=\sigma(x_i,y_j)$ and 
$\delta^x_{i,j}=\sigma_x(x_i,y_j)$ and $\delta^y_{i,j}=\sigma_y(x_i,y_j)$ 
are constant in the cell $D_{i,j}$. 
The error of the discretization is $O((\bar x-x_i)^2+(\bar y-y_j)^2)$.
Equation (\ref{sigmaapro}) is the Taylor expansion in two dimensions.
If a higher order accuracy is required, 
additional terms in the Taylor expansion can be considered. 

Let 
\begin{eqnarray}
\label{eqn:K0}
{\cal K}^{x,0}_{i-i',j-j'}
= \int\!\!\!\int_{D_{i',j'}} 
\frac{(\bar x-x_i)}{\left((\bar x-x_i)^2+(\bar y-y_j)^2\right)^{3/2}}d\bar xd\bar y,
\end{eqnarray}
\begin{eqnarray}
\label{eqn:Kx}
{\cal K}^{x,x}_{i-i',j-j'}
 =  \int\!\!\!\int_{D_{i',j'}} 
\frac{(\bar x-x_i)(\bar x - x_{i'})}{\left((\bar x-x_i)^2+(\bar y-y_j)^2\right)^{3/2}}
d\bar xd\bar y,
\end{eqnarray}
and
\begin{eqnarray}
\label{eqn:Ky}
{\cal K}^{x,y}_{i-i',j-j'}
 =  \int\!\!\!\int_{D_{i',j'}} 
\frac{(\bar x-x_i)(\bar y - y_{j'})}{\left((\bar x-x_i)^2+(\bar y-y_j)^2\right)^{3/2}}
d\bar xd\bar y.
\end{eqnarray}
If the surface density is approximated by (\ref{sigmaapro})
then the force in the $x$-direction defined by (\ref{eqnFxFy}) and (\ref{xforce})
can also be approximated by 
\begin{eqnarray*}
F^x_{i,j} &\approx &\sum^N_{i'=1}\sum^N_{j'=1}
\int\!\!\!\int_{D_{i',j'}} 
\frac{\partial}{\partial x}{\cal K}(\bar x-x_i,\bar y-y_j,0)
\left(
\sigma_{i',j'} 
+\delta^x_{i',j'}(\bar x-x_{i'}) 
+\delta^y_{i',j'}(\bar y-y_{j'})
\right)d\bar x d\bar y\\
&:=& F^{x,0}_{i,j} + F^{x,x}_{i,j} + F^{x,y}_{i,j},
\end{eqnarray*}
where 
\begin{eqnarray}
\label{Fx0ji}
F^{x,0}_{i,j} & = & \sum^N_{i'=1}\sum^N_{j'=1}\sigma_{i',j'}
\int\!\!\!\int_{D_{i',j'}} 
\frac{(\bar x-x_i)}{\left((\bar x-x_i)^2+(\bar y-y_j)^2\right)^{3/2}}d\bar xd\bar y
=\sum^N_{i'=1}\sum^N_{j'=1}\sigma_{i',j'} {\cal K}^{x,0}_{i-i',j-j'},\\
\label{Fxx0ji}
F^{x,x}_{i,j} & = & \sum^N_{i'=1}\sum^N_{j'=1}\delta^x_{i',j'}
\int\!\!\!\int_{D_{i',j'}} 
\frac{(\bar x-x_i)(\bar x - x_{i'})}{\left((\bar x-x_i)^2+(\bar y-y_j)^2\right)^{3/2}}d\bar xd\bar y
=\sum^N_{i'=1}\sum^N_{j'=1}\delta^x_{i',j'} {\cal K}^{x,x}_{i-i',j-j'},\\
\label{Fxy0ji}
F^{x,y}_{i,j} & = & \sum^N_{i'=1}\sum^N_{j'=1}\delta^y_{i',j'}
\int\!\!\!\int_{D_{i',j'}} 
\frac{(\bar x-x_i)(\bar y - y_{j'})}{\left((\bar x-x_i)^2+(\bar y-y_j)^2\right)^{3/2}}d\bar xd\bar y
=\sum^N_{i'=1}\sum^N_{j'=1}\delta^y_{i',j'} {\cal K}^{x,y}_{i-i',j-j'}.
\end{eqnarray}
The evaluation of (\ref{eqn:K0}), (\ref{eqn:Kx}) and (\ref{eqn:Ky})
can be obtained with the help of the following simple integrals,
\begin{eqnarray*}
\int\!\!\!\int \frac{x}{(x^2+y^2)^{3/2}}dxdy  = -\ln (y+\sqrt{x^2+y^2})+C,\quad\quad\>
\int\!\!\!\int \frac{xy}{(x^2+y^2)^{3/2}}dxdy = -\sqrt{x^2+y^2}+C,
\end{eqnarray*}
\begin{eqnarray*}
\int\!\!\!\int \frac{x^2}{(x^2+y^2)^{3/2}}dxdy
= y\ln (x+\sqrt{x^2+y^2})+C, \quad
\int\!\!\!\int \frac{1}{(x^2+y^2)^{3/2}}dxdy
= -\frac{\sqrt{x^2+y^2}}{xy}+C.
\end{eqnarray*}
The value ${\cal K}^{x,0}_{i-i',j-j'}$ is equal to 
\begin{eqnarray}
\label{K0iijj}
{\cal K}^{x,0}_{i-i',j-j'}=
-\ln 
\left(
(\bar y - y_j) +\sqrt{(\bar x-x_i)^2 + (\bar y -y_j)^2}
\right)
\left|^{x_{i'+\frac{1}{2}}}_{x_{i'-\frac{1}{2}}}
\left|^{y_{j'+\frac{1}{2}}}_{y_{j'-\frac{1}{2}}} 
\right.\right.,
\end{eqnarray}
where the notation $\displaystyle g(x)\left|^b_a\right.=g(b)-g(a)$. 
The calculation of ${\cal K}^{x,x}_{i-i',j-j'}$ and ${\cal K}^{x,y}_{i-i',j-j'}$ 
are split into two parts
by the identity $(\bar x-x_i)(\bar x-x_{i'}) = (\bar x-x_i)^2 + (\bar x - x_i)(x_i - x_{i'})$,
and $(\bar x-x_i)(\bar y-y_{j'}) = (\bar x-x_i)(\bar y-y_j) 
+ (\bar x - x_i)(y_j - y_{j'})$, respectively.
It follows that
\begin{eqnarray*}
{\cal K}^{x,x}_{i-i',j-j'}
& = &(x_i-x_{i'}) {\cal K}^{x,0}_{i-i',j-j'}+ \left(
(\bar  y-y_j)\ln (\bar x-x_i+\sqrt{(\bar x-x_i)^2 +(\bar y-y_j)^2)} 
\right)
\left|^{x_{i'+\frac{1}{2}}}_{x_{i'-\frac{1}{2}}}
\left|^{y_{j'+\frac{1}{2}}}_{y_{j'-\frac{1}{2}}}
\right.\right.,\\
{\cal K}^{x,y}_{i-i',j-j'}
& = & (y_j-y_{j'}) {\cal K}^{x,0}_{i-i',j-j'}+\left(
-\sqrt{(\bar x-x_i)^2 + (\bar y-y_j)^2} \right)
\left|^{x_{i'+\frac{1}{2}}}_{x_{i'-\frac{1}{2}}}
\left|^{y_{j'+\frac{1}{2}}}_{y_{j'-\frac{1}{2}}} 
\right.\right..
\end{eqnarray*}

It is worth noting that the form of  
$F^{x,0}_{i,j}$, $F^{x,x}_{i,j}$, and $F^{x,y}_{i,j}$ 
in (\ref{Fx0ji})-(\ref{Fxy0ji}) are a type of convolution. 
It is known that the computational
complexity of a convolution of two vectors can be reduced to $O(N\log_2 N)$ 
with the help of FFT (see Appendix B). 
It implies that the complexity of this method is $O(N^2 (\log_2 N)^2)$.

\section{Self-gravitating force calculation in polar coordinates}
\label{secPolar}

A similar approach is used to develop the method for polar coordinates in this section. 
The singular integral disappears, but the high order of accuracy is not attained 
because there is no explicit form for the integral of an elliptic function.
The method in polar coordinates is described in detail below.
 
The potential function $\Phi$ of (\ref{DPG}) under the assumption $G=1$
in cylindrical coordinate can be expressed as
\begin{eqnarray*}
\Phi(r,\theta,z)=-\int\!\!\!\int\!\!\!\int 
{\cal K}(\bar r,r,\bar \theta,\theta, \bar z-z)\rho(\bar r,\bar \theta,\bar z)\bar rd\bar r d\bar \theta d\bar z,
\end{eqnarray*}
where 
$\displaystyle {\cal K}(\bar r,r,\bar \theta,\theta,z)=\frac{1}{\sqrt{{\bar r}^2-2\bar r r\cos(\bar\theta-\theta)+r^2+z^2}}.$
By (\ref{SigmaDef}), 
the forces on the disk in $r$-direction and $\theta$-direction become
\begin{eqnarray}
\label{rforce}
\frac{\partial}{\partial r} \Phi(r,\theta,0)
=\int\!\!\!\int \frac{\partial}{\partial r}{\cal K}(\bar r,r,\bar \theta,\theta,0)\sigma(\bar r,\bar \theta)
\bar rd\bar r d\bar \theta
\end{eqnarray}
and 
\begin{eqnarray}
\label{thetaforce}
\frac{1}{r}\frac{\partial}{\partial \theta} \Phi(r,\theta,0)
=\frac{1}{r}
\int\!\!\!\int \frac{\partial}{\partial \theta}{\cal K}(\bar r, r,\bar \theta, \theta,0)
\sigma(\bar r,\bar \theta) \bar rd\bar r d\bar \theta.
\end{eqnarray}
We calculate (\ref{rforce}) and (\ref{thetaforce}) by a numerical approach.

Since the support of the surface density is compact, 
contained in a region ${\cal R}=[0,M]\times [0,2\pi]$ for some number $M>0$,
we discretize the radial region in logarithmic form and the azimuthal region uniformly as follows. 
Given a positive integer $N$, we define 
$\Delta \theta=2\pi/N$, $0<\beta_0<1$, $\beta=\beta_0(1-\Delta \theta)$, 
$r_{i+1/2} =\beta^{N-i}M$, $\theta_{j+1/2}=j\Delta\theta$, 
$i,j=0,\ldots,N$,  
$r_i= \frac{1}{2}(r_{i-1/2}+r_{i+1/2})$ and 
$\theta_j= \frac{1}{2}(\theta_{j-1/2}+\theta_{j+1/2})$
where $i,j=1,\ldots,N$. 
It is worth noting that the point $r_i$ should be the center of the cell 
to guarantee the discretization of the surface density is to second order  
and the effect of $\beta_0$ is to refine the mesh. 
Here, the proposed method for polar coordinates is of first order because 
a singular integration occurs (see below).
Furthermore, the region ${\cal R}$ is discretized into the $N^2$ cells,
${\cal R}_{ij}=[r_{i-1/2},r_{i+1/2}]\times [\theta_{j-1/2},\theta_{j+1/2}]$ for $i,j=1,\ldots,N$.
For $j=1,\ldots,N$, the cells ${\cal R}_{1,j}$ do not cover the origin and 
extra cells $\hat {\cal R}_j =[0,r_{1/2}]\times[\theta_{j-1/2},\theta_{j+1/2}]$
should be included. For simplification of notation,
we denote ${\cal R}_{0,j}=\hat{\cal R}_j$, $j=1,\ldots,N$.

The forces in the $r$-direction and the $\theta$-direction at the point $(r_i,\theta_j)$
of the cell ${\cal R}_{ij}$ are 
\begin{eqnarray}
F^r_{i,j}=\frac{\partial}{\partial r}\Phi(r_i,\theta_j,0),\mbox{\quad and \quad}
F^\theta_{i,j}=\frac{1}{r_i}\frac{\partial}{\partial \theta}\Phi(r_i,\theta_j,0).
\end{eqnarray}
The surface density 
$\sigma$ on $R_{i,j}$ in (\ref{rforce}) is linearly approximated by 
\begin{eqnarray}
\label{sigmaapro2}
\sigma(\bar r,\bar \theta)\approx \sigma_{i,j} 
+\delta^r_{i,j}(\bar r-r_i) + \delta^\theta_{i,j}(\bar \theta-\theta_j),
\end{eqnarray}
where $\sigma_{i,j}=\sigma(r_i,\theta_j)$ and 
$\delta^r_{i,j}=\sigma_r(r_i,\theta_j)$ 
and $\delta^\theta_{i,j}=\sigma_\theta(r_i,\theta_j)$ are constant in the cell $R_{i,j}$. 
The error of the discretization is $O((\bar r-r_i)^2+(\bar \theta-\theta_j)^2)$.
Equation (\ref{sigmaapro2}) is the Taylor expansion in two dimensions.

\subsection{The calculation of radial forces}
Let 
\begin{eqnarray}
\label{eqn:Kr0}
{\cal K}^{r,0}_{i-i',j-j'}
= \int\!\!\!\int_{R_{i',j'}} 
\frac{\bar r(r_i-\bar r\cos({\bar \theta-\theta_j}))}{\left({\bar r}^2+r^2_i - 2\bar r r_i \cos (\bar \theta-\theta_j)\right)^{3/2}}
d\bar rd\bar \theta,
\end{eqnarray}
\begin{eqnarray}
\label{eqn:Krr}
r_i{\cal K}^{r,r}_{i-i',j-j'}
 = \int\!\!\!\int_{R_{i',j'}} 
\frac{\bar r(r_i-\bar r\cos(\bar\theta-\theta_j))(\bar r - r_{i'})}{\left({\bar r}^2+r^2_i - 2\bar r r_i \cos (\bar \theta-\theta_j)\right)^{3/2}}
d\bar rd\bar \theta,
\end{eqnarray}
and
\begin{eqnarray}
\label{eqn:Krt}
{\cal K}^{r,\theta}_{i-i',j-j'}
 = \int\!\!\!\int_{R_{i',j'}} 
\frac{\bar r(r_i-\bar r\cos(\bar\theta-\theta_j))(\bar \theta - \theta_{j'})}
{\left({\bar r}^2+r^2_i - 2\bar r r_i \cos (\bar \theta-\theta_j)\right)^{3/2}}
d\bar rd\bar \theta.
\end{eqnarray}
The term $r_{i}$ in (\ref{eqn:Krr}) is for the formulation of
a convolution type. 
By (\ref{rforce}) and (\ref{sigmaapro2}), we have 
\begin{eqnarray*}
F^r_{i,j} &\approx &\sum^N_{i'=0}\sum^N_{j'=1}
\int\!\!\!\int_{R_{i',j'}} 
\frac{\partial}{\partial r}{\cal K}(\bar r,r_i,\bar \theta,\theta_j,0)
\left(
\sigma_{i',j'} 
+\delta^r_{i',j'}(\bar r-r_{i'}) 
+\delta^\theta_{i',j'}(\bar \theta-\theta_{j'})
\right)\bar rd\bar r d\bar \theta\\
&:=& F^{r,0}_{i,j} + F^{r,r}_{i,j} + F^{r,\theta}_{i,j},
\end{eqnarray*}
where 
\begin{eqnarray}
\label{Fr0ji}
F^{r,0}_{i,j}  & = & \sum^N_{i'=0}\sum^N_{j'=1}\sigma_{i',j'}
\int\!\!\!\int_{R_{i',j'}} 
\frac{\bar r(r_i-\bar r\cos(\bar\theta-\theta_j))}
{\left({\bar r}^2+r^2_i - 2\bar r r_i \cos (\bar \theta-\theta_j)\right)^{3/2}}
d\bar rd\bar \theta\\
\label{Frr0ji}
F^{r,r}_{i,j}  & = & \sum^N_{i'=0}\sum^N_{j'=1}\delta^r_{i',j'}
\int\!\!\!\int_{R_{i',j'}} 
\frac{\bar r(r_i-\bar r\cos(\bar\theta-\theta_j)(\bar r-r_{i'})}
{\left({\bar r}^2+r^2_i - 2\bar r r_i \cos (\bar \theta-\theta_j)\right)^{3/2}}
d\bar rd\bar \theta \\
\label{Frt0ji}
F^{r,\theta}_{i,j} & = & \sum^N_{i'=0}\sum^N_{j'=1}\delta^\theta_{i',j'}
\int\!\!\!\int_{R_{i',j'}} 
\frac{\bar r(r_i-\bar r\cos(\bar\theta-\theta_j))(\bar \theta - \theta_{j'})}
{\left({\bar r}^2+r^2_i - 2\bar r r_i \cos (\bar \theta-\theta_j)\right)^{3/2}}
d\bar rd\bar \theta
\end{eqnarray}
Equations (\ref{Fr0ji}), (\ref{Frr0ji}), and (\ref{Frt0ji}) can be rewritten as 
\begin{eqnarray}
\label{Fr0ji2}
F^{r,0}_{i,j} &= &\sum^N_{i'=1}\sum^N_{j'=1}\sigma_{i',j'} {\cal K}^{r,0}_{i-i',j-j'}
+\sum^N_{j'=1}\sigma_{0,j'}{\bar{\cal K}}^{r,0}_{i,j-j'},\\
\label{Frr0ji2}
F^{r,r}_{i,j}&=&r_i\sum^N_{i'=1}\sum^N_{j'=1}\delta^r_{i',j'} {\cal K}^{r,r}_{i-i,j-j'}
             +r_i\sum^N_{j'=1}\delta^r_{0,j'}{\bar {\cal K}}^{r,r}_{i,j-j'},\\
\label{Frt0ji2}
F^{r,\theta}_{i,j} &=&\sum^N_{i'=1}\sum^N_{j'=1}\delta^\theta_{i',j'} {\cal K}^{r,\theta}_{i-i',j-j'}
                   +\sum^N_{j'=1}\delta^\theta_{0,j'}\bar{\cal K}^{r,\theta}_{i,j-j'}.
\end{eqnarray}

Let us define $F(\tilde r,\theta)=\sqrt{1+{\tilde r}^2-2{\tilde r}\cos(\theta)}$,
where $\tilde r$ is a dimensionless radius.
The evaluation of (\ref{eqn:Kr0}), (\ref{eqn:Krr}) and (\ref{eqn:Krt})
can be obtained with the help of the following simple integrals,
\begin{eqnarray*}
\int \frac{\bar r(r-\bar r\cos(\theta))}{\left({\bar r}^2+r^2-2 r\bar r\cos(\theta)\right)^{3/2}}d\bar r  
&=& 
-\cos(\theta)\ln (-\cos(\theta)+\frac{\bar r}{r}+F(\frac{\bar r}{r},\theta))
+\frac{2\cos(\theta)\frac{\bar r}{r}-1}{F(\frac{\bar r}{r},\theta)}+C\\
&:=& H_1(\frac{\bar r}{r},\theta)+C
\end{eqnarray*}
and
\begin{eqnarray*}
\int \frac{{\bar r}^2(r-\bar r\cos(\theta))}{\left({\bar r}^2+r^2-2 r\bar r\cos(\theta)\right)^{3/2}}d\bar r 
& = &
-r\left(
(3\cos^2(\theta)-1)
\ln (-\cos(\theta)+\frac{\bar r}{r}+F(\frac{\bar r}{r},\theta))\right. \\
&+&\left.\frac{1}{F(\frac{\bar r}{r},\theta)}
(-6\frac{\bar r}{r}\cos^2(\theta)+3\cos(\theta)+\frac{{\bar r}^2}{r^2}\cos(\theta)
+\frac{\bar r}{r})\right)+C\\
&:=&r H_2(\frac{\bar r}{r},\theta)+C.
\end{eqnarray*}

Following the definition of $r_{i'+1/2}$ and $r_i$, we have 
\begin{eqnarray*}
\frac{r_{i'+1/2}}{r_i}=\frac{2\beta^{i-i'}}{1+\beta},\mbox{ and }
\frac{r_{i'}}{r_i} = \beta^{i-i'}.
\end{eqnarray*}
We calculate the value of the integral
\begin{eqnarray*}
{\cal K}^{r,0}_{i-i',j-j'}
&=&
\int^{\theta_{j'+1/2}}_{\theta_{j'-1/2}}
\!\!\! 
\int^{r_{i'+1/2}}_{r_{i'-1/2}} 
\frac{\bar r(r_i-\bar r\cos(\bar\theta-\theta_j))}{\left({\bar r}^2+r^2_i - 2\bar r r_i \cos (\bar \theta-\theta_j)\right)^{3/2}}
d\bar rd\bar \theta\\
&=&\int^{\theta_{j'+1/2}}_{\theta_{j'-1/2}}
-\cos(\bar\theta-\theta_j)\ln (-\cos(\bar\theta-\theta_j)+\bar r/r_i+F(\bar r/r_i,\bar\theta-\theta_j))\\
&+&\frac{2\cos(\bar\theta-\theta_j)\bar r/r_i-1}{F(\bar r/r_i,\bar\theta-\theta_j)}
\left|^{r_{i'+1/2}}_{r_{i'-1/2}}\right.
d\bar\theta
\end{eqnarray*}
The last integral in the above equation is an elliptic integral and a 
trapzoidal rule has been applied for its evaluation. 
It is of second order accuracy for the integration of a smooth function.
Unfortunately, the presence of a singular function in terms of $\ln(1-\cos(\theta))$
reduces the accuracy of the proposed method for polar coordinate to first order. 
Finally, the value ${\cal K}^{r,0}_{i-i',j-j'}$ is approximated as follows and is used in the numerical 
calculation,
\begin{eqnarray*}
{\cal K}^{r,0}_{i-i',j-j'}
&\approx&
\frac{1}{2}\left( H_1(r_{i'+1/2}/r_i,\theta_{j'+1/2}-\theta_j)
\quad -H_1(r_{i'-1/2}/r_i,\theta_{j'+1/2}-\theta_j)\right.\\
&+&H_1(r_{i'+1/2}/r_i,\theta_{j'-1/2}-\theta_j)
-H_1(r_{i'-1/2}/r_i,\theta_{j'-1/2}-\theta_j)\left.\right)(\theta_{j'+1/2}-\theta_{j'-1/2})\\
&:=&H_1(\frac{\bar r}{r_i},{\bar\theta-\theta_j})
\left|^{r_{i'+1/2}}_{r_{i'-1/2}}\right.
\left]^{\theta_{j'+1/2}}_{\theta_{j'-1/2}}\right.,
\end{eqnarray*} 
where the notation $f(\cdot)]^b_a = \frac{1}{2}(f(b)+f(a))(b-a)$.
Similarly, 
\begin{eqnarray*}
{\cal K}^{r,r}_{i-i',j-j'}
&\approx &
H_2(\frac{\bar r}{r_i},{\bar \theta-\theta_j})
\left|^{r_{i'+1/2}}_{r_{i'-1/2}}\right.
\left]^{\theta_{j'+1/2}}_{\theta_{j'-1/2}}\right.-\frac{r_{i'}}{r_i}{\cal K}^{r,0}_{i-i',j-j'},\\
{\cal K}^{r,\theta}_{i-i',j-j'}
&\approx&
(\bar\theta-\theta_j) H_1(\frac{\bar r}{r_i},\bar\theta-\theta_j)
\left|^{r_{i'+1/2}}_{r_{i'-1/2}}\right.
\left]^{\theta_{j'+1/2}}_{\theta_{j'-1/2}}\right..\\
\end{eqnarray*}

\subsection{The calculation of azimuthal forces}
Next, we introduce the calculation for ${\cal K}^{\theta,0}_{i-i',j-j'}$, 
${\cal K}^{\theta,r}_{i-i',j-j'}$, and 
${\cal K}^{\theta,\theta}_{i-i',j-j'}$.

\noindent
In particular, we calculate the value of the integral
\begin{eqnarray*}
{\cal K}^{\theta,0}_{i-i',j-j'}
&=&r_i
\int^{\theta_{j'+1/2}}_{\theta_{j'-1/2}}
\!\!\! 
\int^{r_{i'+1/2}}_{r_{i'-1/2}} 
\frac{{\bar r}^2 \sin({\bar \theta-\theta_j})}{\left({\bar r}^2+r^2_i - 2\bar r r_i \cos (\bar \theta-\theta_j)\right)^{3/2}}
d\bar rd\bar \theta\\
&=& - r_i \left(
F(\frac{\bar r}{r_i},{\bar \theta-\theta_j})+\frac{\bar r}{r_i}\ln(-\cos(\bar \theta-\theta_j)+\frac{\bar r}{r_i}
+F(\frac{\bar r}{r_i},{\bar \theta-\theta_j}))
\right)\left|^{r_{i'+1/2}}_{r_{i'-1/2}}\!\left|^{\theta_{j'+1/2}}_{\theta_{j'-1/2}}\right.\right..
\end{eqnarray*} 
Similarly, 
\begin{eqnarray*}
{\cal K}^{\theta,r}_{i-i',j-j'}
&=&
- r_i
(-1+\frac{\bar r}{2r_i}+\frac{3}{2}\cos(\bar\theta-\theta_j))
F(\frac{\bar r}{r_i},{\bar \theta-\theta_j}) \\
&-& r^2_i
(\frac{3}{2}\cos^2({\bar \theta-\theta_j})
-\frac{1}{2}-\cos({\bar \theta-\theta_j}) )
\ln(-\cos(\bar \theta-\theta_j)+\frac{\bar r}{r_i}
+F(\frac{\bar r}{r_i},{\bar \theta-\theta_j}))
\left|^{r_{i'+1/2}}_{r_{i'-1/2}}\!\left|^{\theta_{j'+1/2}}_{\theta_{j'-1/2}}\right.\right..
\end{eqnarray*}
and 
\begin{eqnarray*}
{\cal K}^{\theta,\theta}_{i-i',j-j'}
\approx
r_i\frac{{\bar \theta}-\theta_j}{-1+\cos^2({\bar\theta-\theta_j})}
\left(
\sin({\bar \theta}-\theta_j)
(\frac{\bar r}{r_i}-2\cos^2({\bar \theta}-\theta_j)\frac{\bar r}{r_i}
+\cos({\bar \theta}-\theta_j))\right.\\
\left.
+(\cos^2({\bar \theta}-\theta_j)-\sin({\bar \theta}-\theta_j))
\ln(-\cos({\bar \theta}-\theta_j)+\frac{\bar r}{r_i}+F(\frac{\bar r}{r_i},{\bar\theta-\theta_j})
\right)\left|^{r_{i'+1/2}}_{r_{i'-1/2}}\right.\left]^{\theta_{j'+1/2}}_{\theta_{j'-1/2}}\right.
\end{eqnarray*}

\section{Order of accuracy and a comparison study}
\label{secAccurcy}
\subsection{Order of accuracy}
We investigate the numerical errors induced by the truncation introduced in 
(\ref{sigmaapro}), which is 
\begin{eqnarray*}
O\left(((\Delta x)^2+(\Delta y)^2)
\int\!\!\!\int_{D_{i',j'}} 
\frac{|\bar x-x_i|}{\left((\bar x-x_i)^2+(\bar y-y_j)^2\right)^{3/2}}d\bar x d\bar y
\right).
\end{eqnarray*}
The last integral in the above estimation is $O(|\ln \Delta x|)$
which gives the numerical error of order $O((\Delta x)^2 |\ln \Delta x|)=O((\Delta x)^{2-})$
with $\Delta x=\Delta y$. 
Three types of norm are used to measure the errors between the numerical
and analytic solutions. The 
$L^1$, $L^2$, and $L^\infty$ norms  of a function $f$ on a domain $\Omega$ are defined as 
\begin{eqnarray*}
\|f\|_p=\left(\int_\Omega |f(\vec{x})|^p d\vec{x}\right)^{1/p},\mbox{ for } p= 1,2,
\mbox{ and } 
\|f\|_\infty=\mbox{ess sup}_{\Omega}|f(\vec{x})|.
\end{eqnarray*}
The errors between the analytic and numerical solutions for 
various resolutions using different norms $L^1$, $L^2$,
and $L^\infty$ demonstrate the convergence in 
total variation, energy, and pointwise senses, respectively.

We verify that the proposed method is of second order accuracy 
by demonstrating the following examples, 
a $D_2$ disk~\cite{ESchulz},  a non-axisymmetric disk consisting of two superposed $D_2$ disks
and a non-axisymmetric disk describing a pair of spirals.

\begin{figure}
\begin{center}
\includegraphics[width=.43\textwidth]{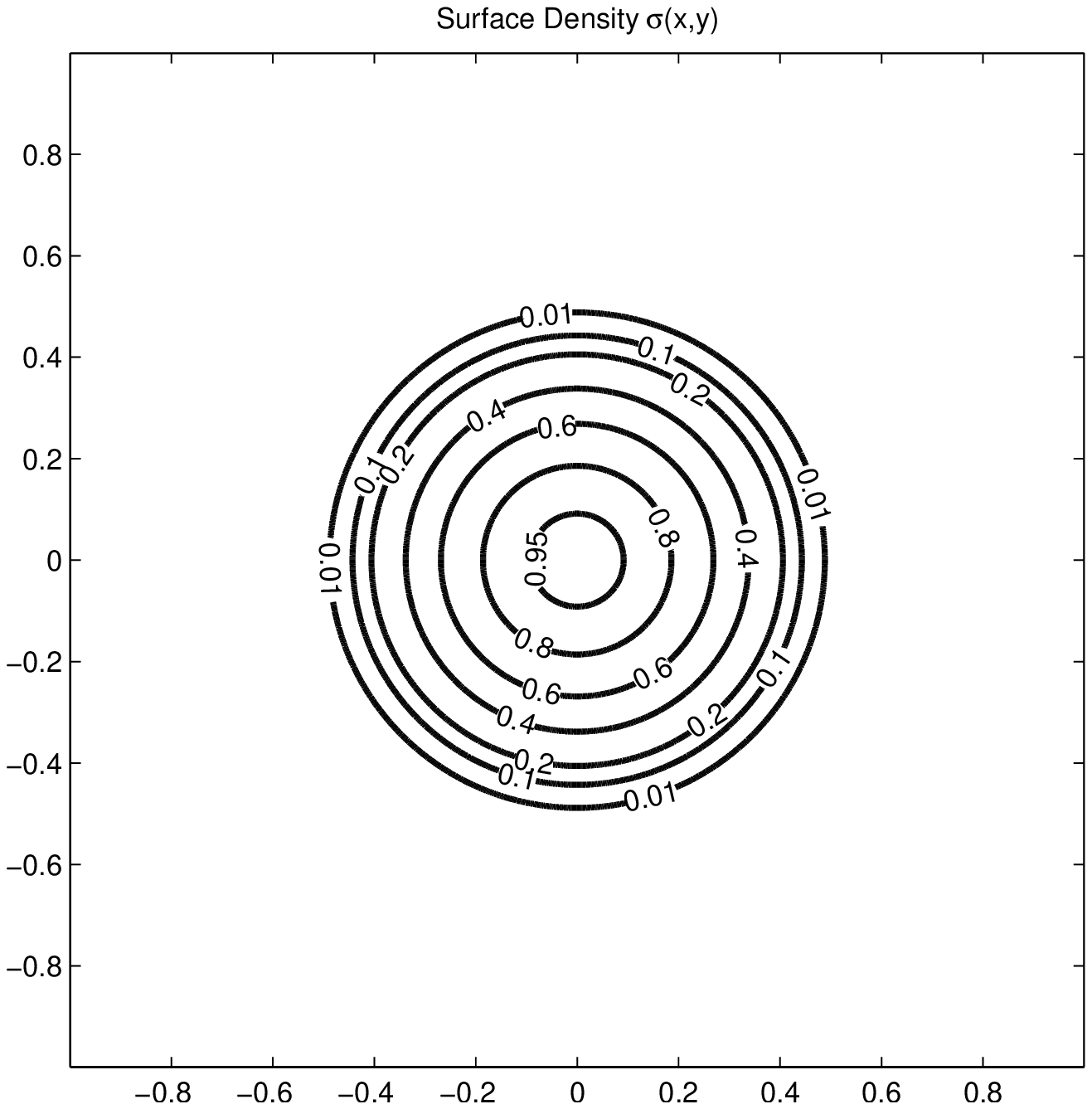}
\includegraphics[width=.43\textwidth]{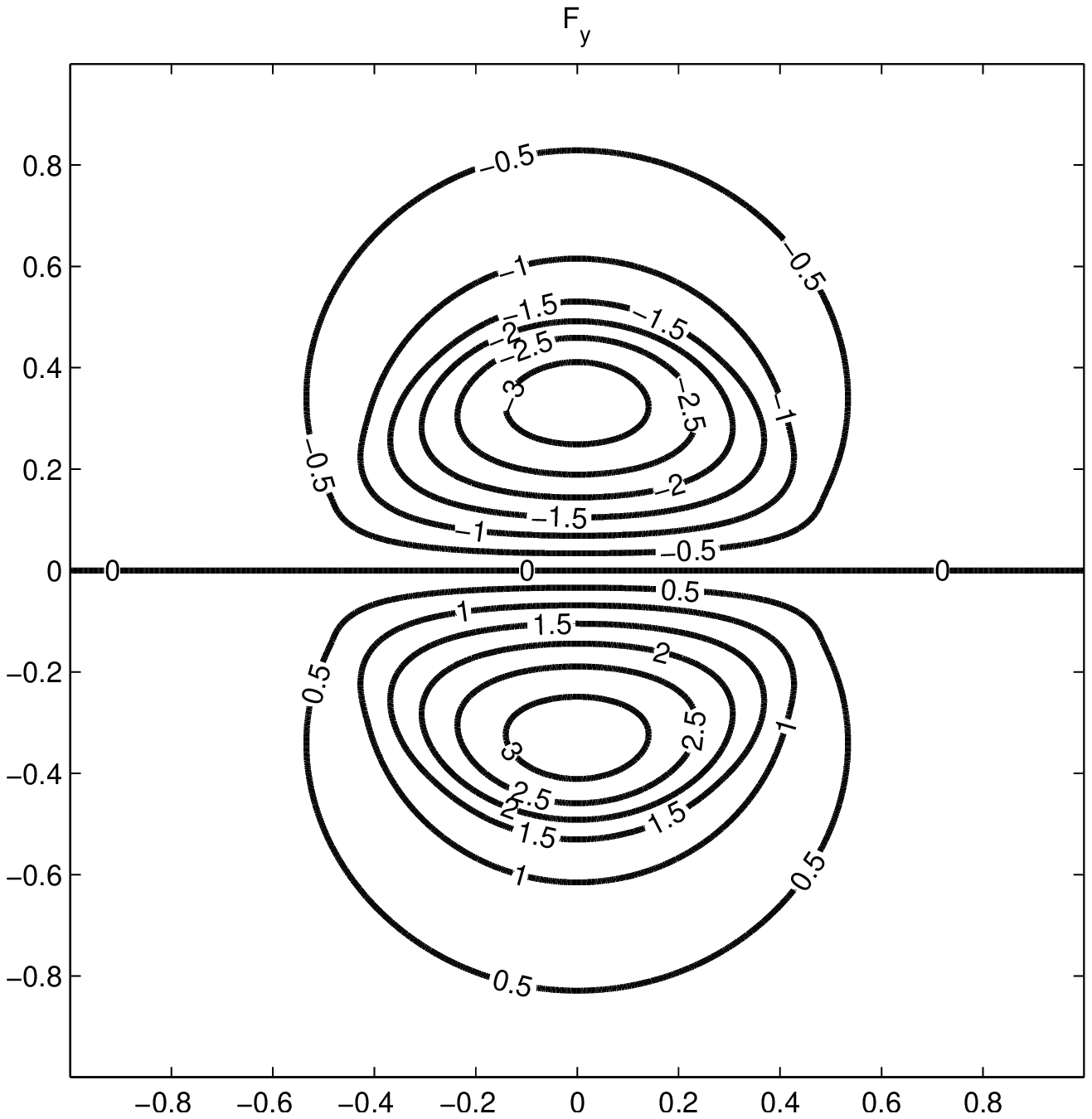}
\end{center}
\begin{center}
\includegraphics[width=.43\textwidth]{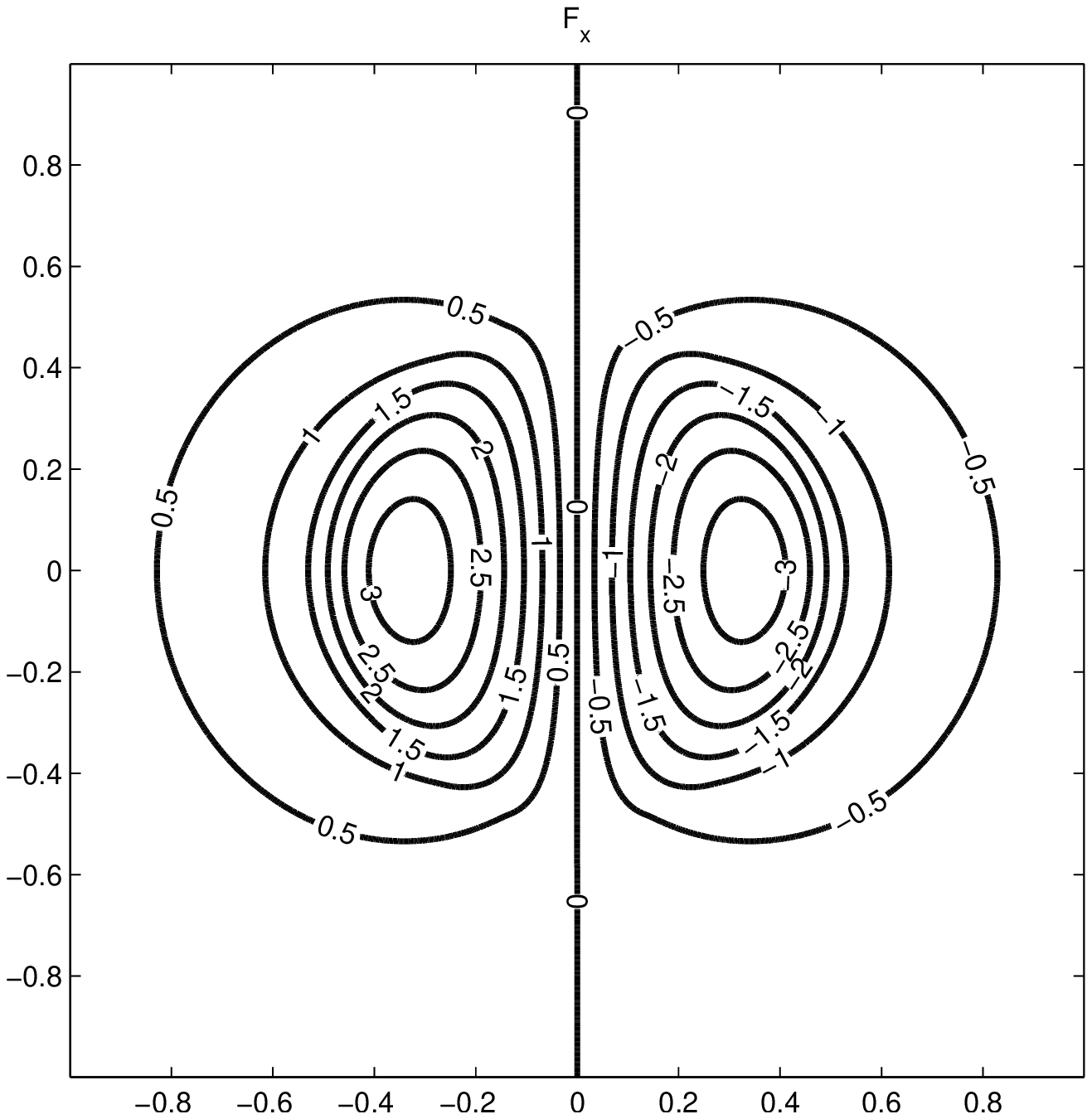}
\includegraphics[width=.43\textwidth]{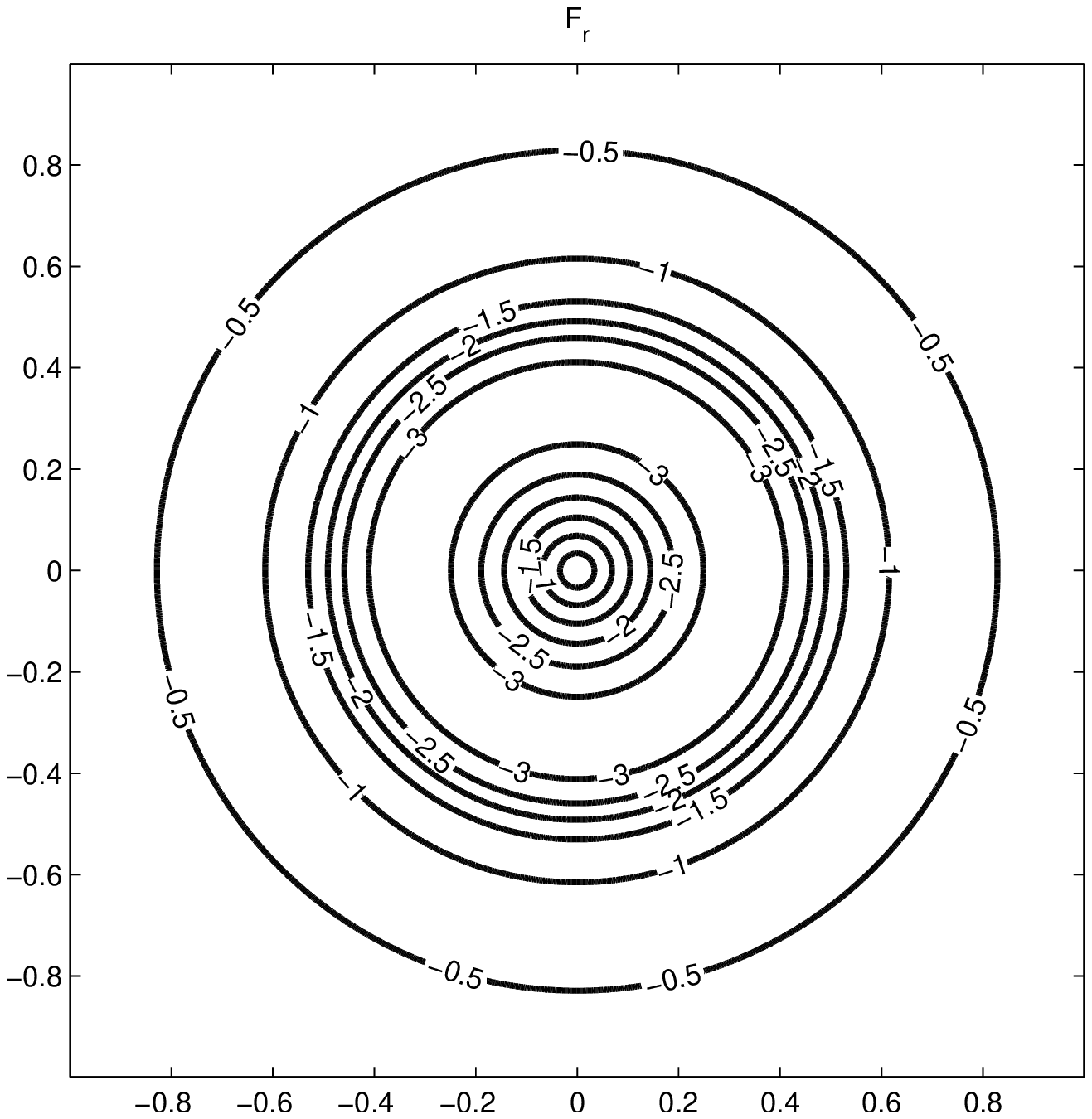}
\end{center}
\begin{center}
\includegraphics[width=.43\textwidth]{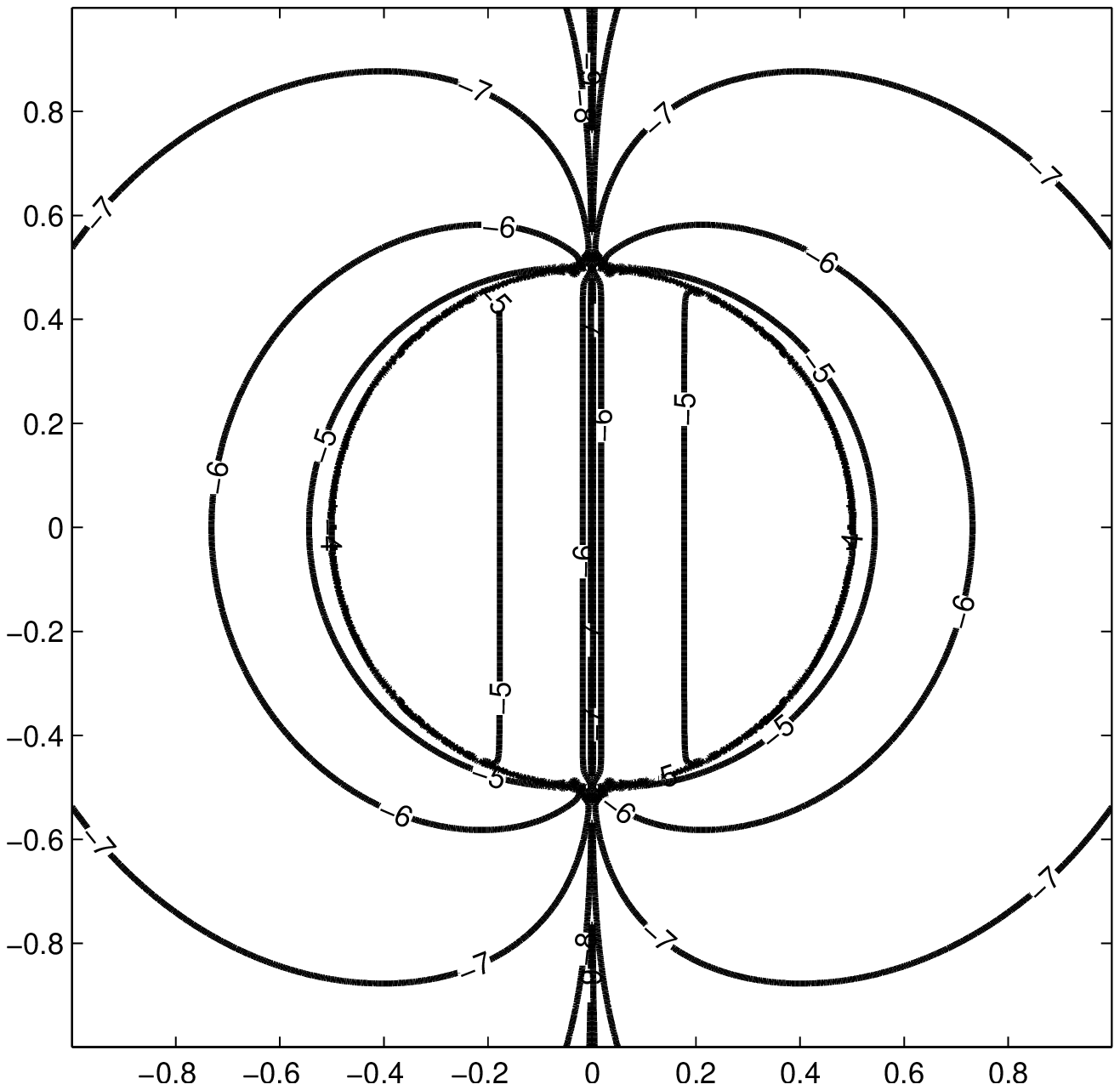}
\includegraphics[width=.43\textwidth]{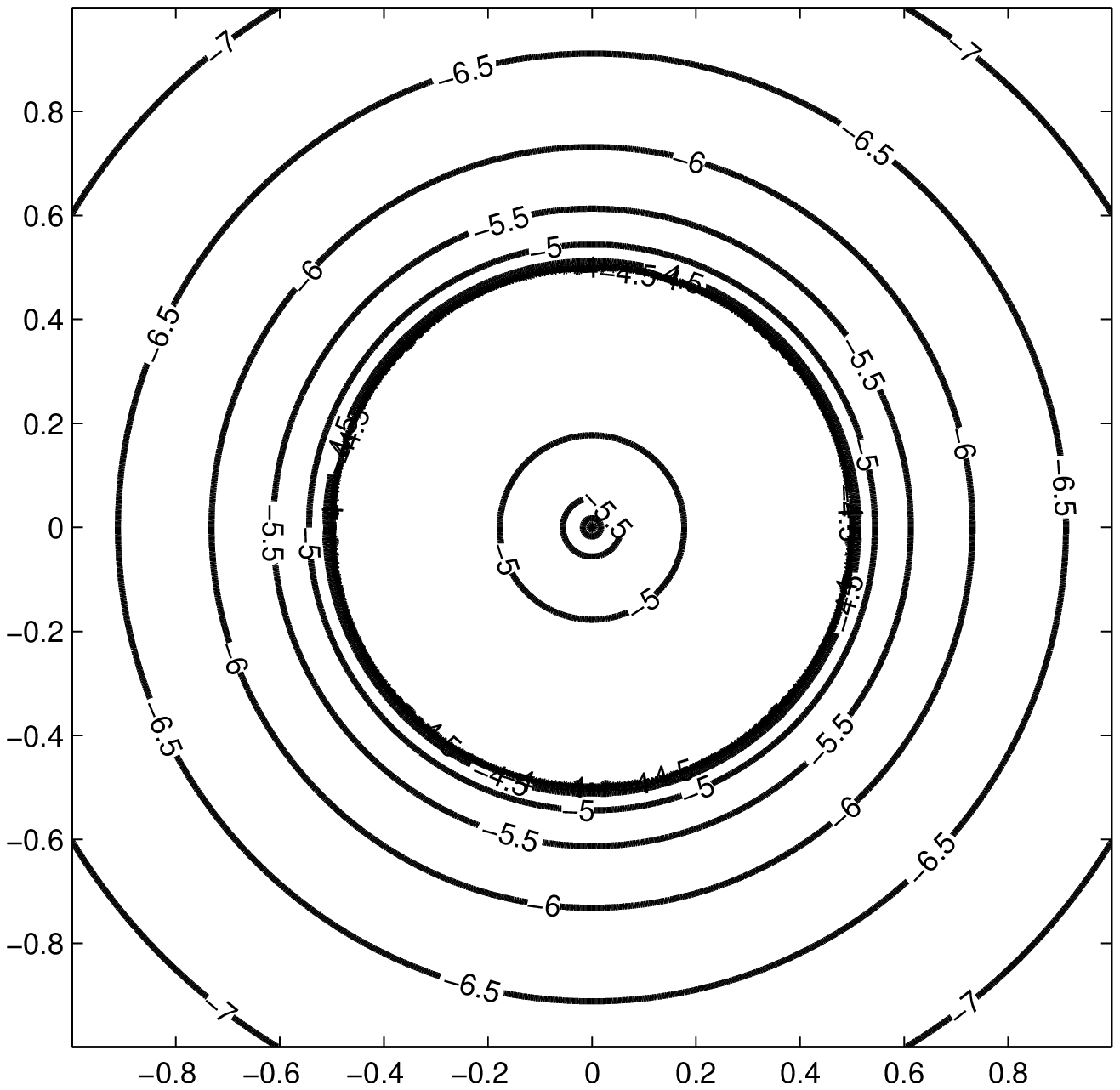}
\end{center}
\caption{The numerical solutions of a $D_2$ disk for $N=1024$,
the contour plots are surface density (upper left), 
the $y$-directional force (upper right),
the $x$-directional force (middle left), 
the radial force (middle right),
the difference between analytic and numerical solutions
in $x$ direction (lower left), 
and the difference in radial direction (lower right).
The values in the lower contour plots are 
the absolute difference in the common logarithmic scale. 
}
\label{D2N1024}
\end{figure}

\vskip 0.5cm
\noindent{\it Example 1}.
The $D_2$ disk has the surface density  
\begin{eqnarray}
\Sigma_{D_2}(R;\alpha) 
=\left\{
\begin{array}{ll}
\sigma_0 (1-R/\alpha^2)^{3/2} & \mbox{ for } R<\alpha,\\
0 & \mbox{ for } R>\alpha,
\end{array}
\right.
\end{eqnarray}
where $R=\sqrt{x^2+y^2}$ and $\alpha$ is a given constant.
The corresponding potential on the $z=0$ plane is 
\begin{eqnarray*}
\Phi_{D_2}(R,0;\alpha) 
=\left\{
\begin{array}{ll}
-\frac{3\pi^2\sigma_0 R G}{64\alpha^3}(8\alpha^4-8\alpha^2R^2+3R^4) & \mbox{ for } R\le \alpha\\
-\frac{3\pi\sigma_0 G}{32\alpha}
\left[(8\alpha^4-8\alpha^2R^2+3R^4)\sin^{-1}(\frac{\alpha}{R})
+3\alpha(2\alpha^2-R^2)\sqrt{R^2-\alpha^2}\right] & \mbox{ for } R\ge \alpha,
\end{array}
\right.
\end{eqnarray*}
and the radial force is found as 
\begin{eqnarray*}
F_{R,D_2}(R,0;\alpha) 
=\left\{
\begin{array}{ll}
-\frac{3\pi^2\sigma_0 R G}{16\alpha^3}(4\alpha^2-3R^2) & \mbox{ for } R\le \alpha\\
-\frac{3\pi\sigma_0 G}{8\alpha^3}\left[R(4\alpha^2-3R^2)\sin^{-1}(\frac{\alpha}{R})
-\alpha(2\alpha^2-3R^2)\sqrt{1-\alpha^2/R^2}\right] & \mbox{ for } R\ge \alpha.
\end{array}
\right.
\end{eqnarray*} 
Without loss of generality for studying the order of accuracy,
let us set the computational domain $\Omega=[-1,1]\times [-1,1]$,
$\sigma_0=G=1$ and $\alpha=0.25$. 
We illustrate the contour plots
of the surface density, $x$-directional force, $y$-directional force,
radial force, residuals between analytic and numerical solutions
for $x$ , and radial directions for $N=1024$ in Fig.~\ref{D2N1024}. 
The residuals show that the largest errors occur in regions 
surrounding the edge of the disk where the second derivative
of the surface density with respect to radius 
is infinite.

In Table~\ref{tblD2},
the column $E^p_x$ and $E^p_R$ is the error of
the $x$ directional force and $R$ radial direction by $p$-norm, $p=1,2$, 
and $\infty$, between the analytic and numerical solutions.
The column $O^p_x$ in Table~\ref{tblD2} is the order of accuracy as measured 
by $\log_2(E^p_x(2^{k-1})/E^p_x(2^k))$ for $k=6$ to $10$ and similarly for $O^p_R$.
These errors show that this method is nearly of second order accuracy. 
More precisely, we obtain the order of convergence to be about $1.8$ or $1.9$
as measured by the $L^1$ and $L^2$ norms for the simulation of a $D_2$ disk.
The $L^\infty$ norm only demonstrates the convergence,
since the second derivatives of the surface density of the $D_2$ disk are not bounded. 

\vskip 0.5cm

\begin{table}
\begin{center}
\begin{tabular}{|c|c|c|c||c|c|c|}  \hline
$N$ & $E^1_x$   & $E^2_x$   & $E^\infty_x$  & $E^1_R$   &  $E^2_R$  & $L^\infty_R$ \\ \hline 
32	&1.156E-2	&1.134E-2	&2.231E-2	    &1.788E-2   &1.589E-2	&2.315E-2 \\ \hline
64	&3.039E-3	&3.176E-3	&7.525E-3	    &4.742E-3	&4.460E-3	&8.535E-3 \\ \hline
128	&8.476E-4	&9.312E-4	&5.264E-3	    &1.319E-3	&1.309E-3	&5.906E-3 \\ \hline
256	&2.161E-4	&2.444E-4	&1.932E-3	    &3.379E-4	&3.439E-4	&1.994E-3 \\ \hline
512	&5.620E-5	&6.884E-5	&9.478E-4	    &8.795E-5	&9.695E-5	&9.842E-4 \\ \hline
1024&1.427E-5	&1.824E-5	&3.281E-4	    &2.236E-5	&2.570E-5	&3.470E-4 \\ \hline \hline
$N_{k-1}/N_k$ &   $O^1_x$   & $O^2_x$   & $O^\infty_x$  & $O^1_R$   &  $O^2_R$  & $O^\infty_R$ \\ \hline
32/64	&1.927	&1.836	&1.567	&1.914	&1.833	&1.439 \\ \hline
64/128	&1.842	&1.770	&0.515	&1.846	&1.768	&0.531 \\ \hline
128/256	&1.971	&1.929	&1.446	&1.964	&1.928	&1.566 \\ \hline
256/512	&1.943	&1.827	&1.027	&1.941	&1.826	&1.018 \\ \hline
512/1024&1.977	&1.916	&1.530	&1.975	&1.915	&1.504 \\ \hline
\end{tabular}
\end{center}
\caption{This table demonstrates
the errors and order accuracy of the proposed method 
for the $D_2$ disk for various number of zones $N=2^k$ from
$k=5$ to $10$.  
It shows that the order for the $D_2$ disk is about $1.8$ or
$1.9$ order in $L^1$ and $L^2$ norm.  
}
\label{tblD2}
\end{table}

\begin{figure}
\begin{center}
\includegraphics[width=.31\textwidth]{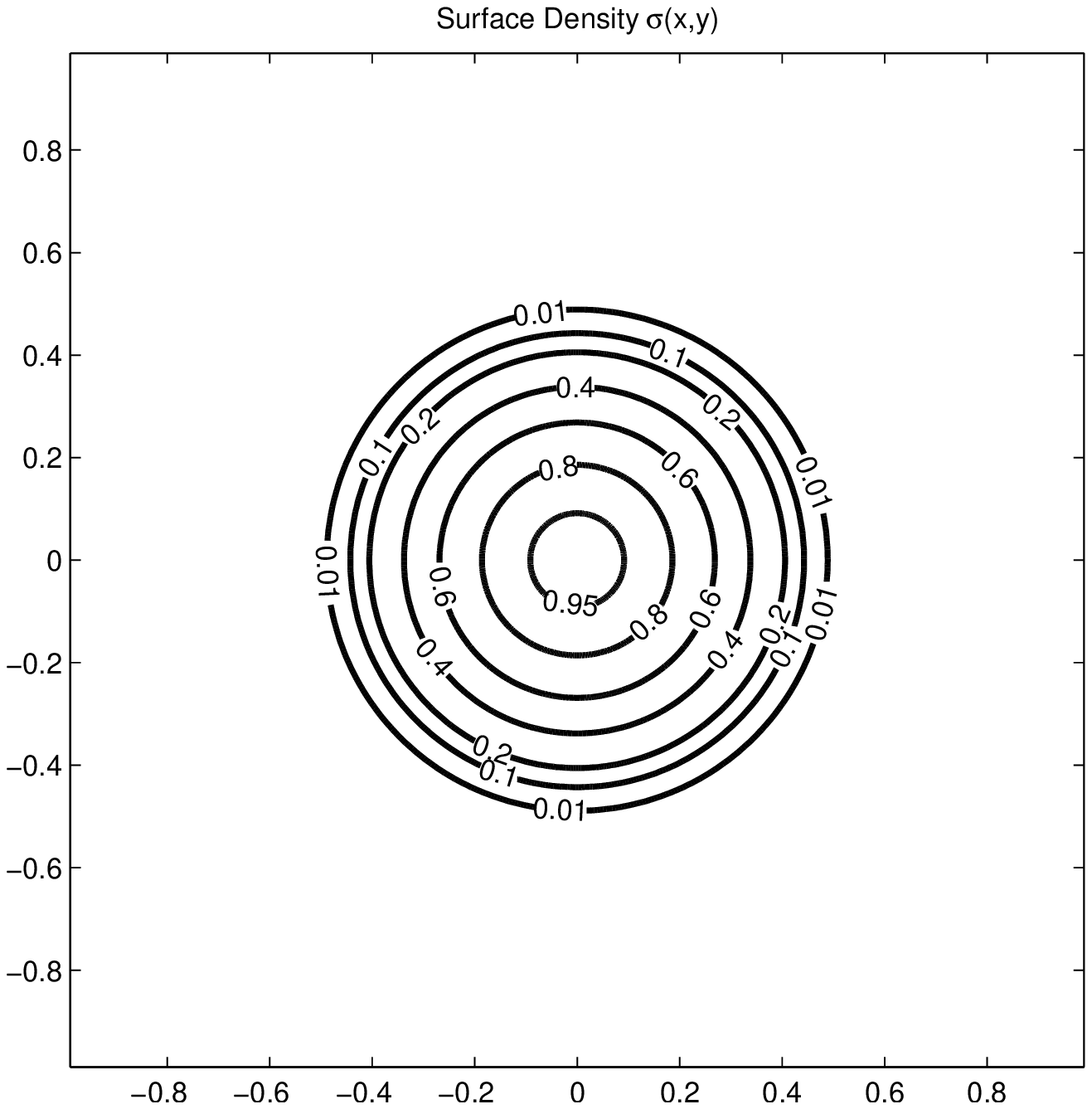}
\includegraphics[width=.31\textwidth]{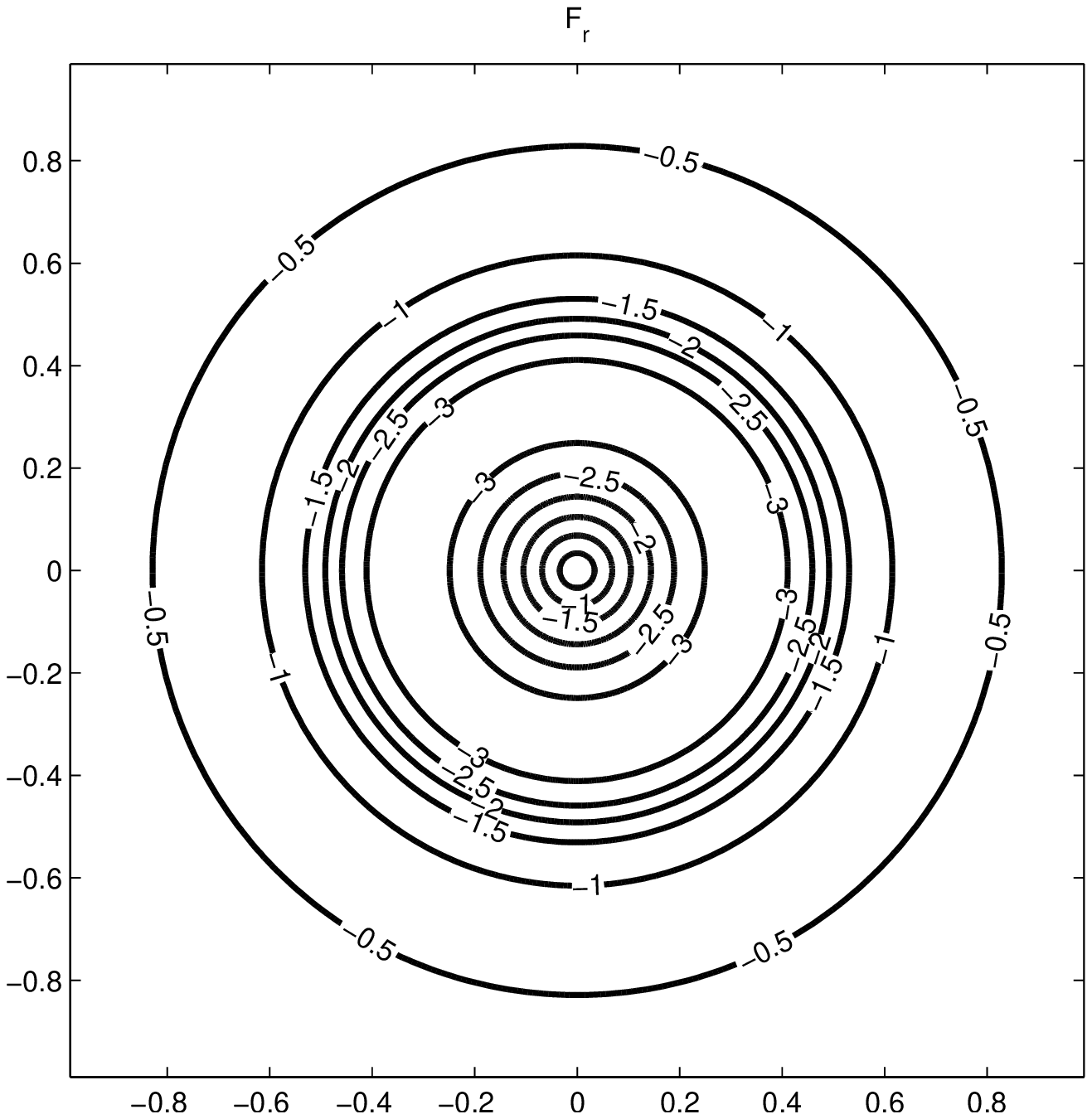}
\includegraphics[width=.31\textwidth]{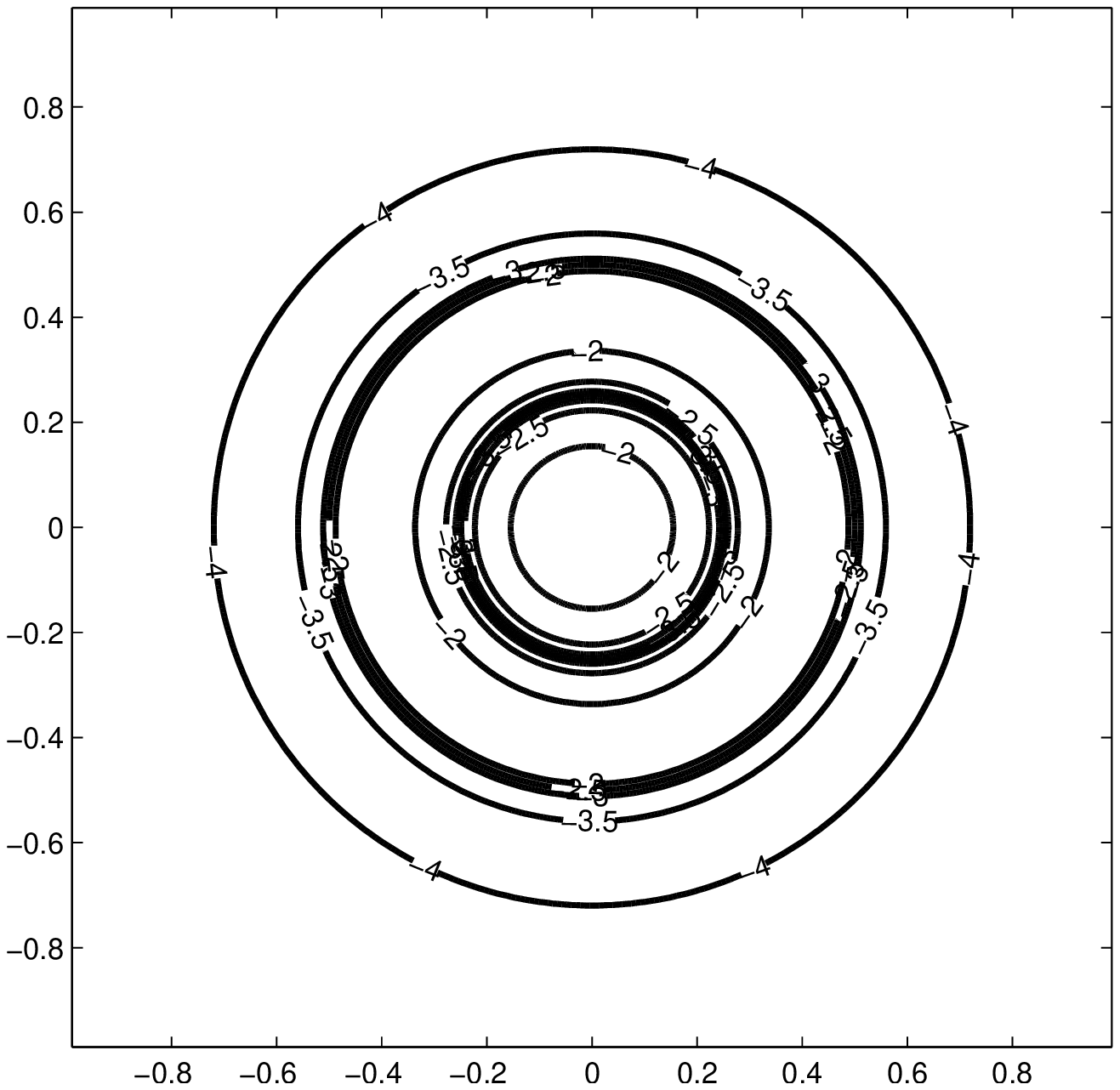}
\end{center}
\caption{The numerical solutions of a $D_2$ disk for $N=512$
to investigate the self-gravitational force calculation in polar coordinate.
From left to right, 
the contour plots are surface density, 
the radial directional force,
the difference between analytic and numerical solutions, respectively.
The values in the right contour plot are 
the absolute difference in the common logarithmic scale. 
}
\label{D2N512P}
\end{figure}

We continue to  use the $D_2$ disk as an example 
and a unit disk  $D(0,1)=\Omega=[0,1]\times [0,2\pi]$ 
as the computational domain to investigate the self-gravitational force in
polar coordinates. The value $\beta_0=0.99$ is set. We show the contour plots
of the surface density, radial force, and the difference between
analytic and numerical solutions for $N=512$ in Fig.~\ref{D2N512P}
and the order of accuracy is only about 1 as given in Table~\ref{tblPD2}. 
The largest errors occur in regions not only surrounding the edge of the disk, 
but also close to the origin. Although the surface density
at the origin is smooth, the singular elliptic integral introduces significant error there. 
Hereafter, we concentrate on the self-gravitational forces in Cartesian coordinates.

\begin{table}
\begin{center}
\begin{tabular}{|c|c|c|c||c|c|c|c|}  \hline
$N$ & $E^1_R$   &  $E^2_R$  & $L^\infty_R$ & $N_{k-1}/N_k$ &  $O^1_R$ & $O^2_R$ & $O^\infty_R$ \\ \hline 
32	&1.603E-1   &1.725E-1	&2.725E-1      &               &          &         &               \\ \hline
64	&7.618E-2	&8.289E-2	&1.361E-1      & 32/64         & 1.073    & 1.057   &  1.002        \\ \hline
128	&3.646E-2	&4.045E-2	&6.806E-2      & 64/128        & 1.063    & 1.035   &  1.000        \\ \hline
256	&1.754E-2	&2.098E-2	&3.403E-2      & 128/256       & 1.056    & 0.947   &  1.000        \\ \hline
512	&8.762E-3	&1.049E-2	&1.701E-2      & 256/512       & 1.001    & 1.000   &  1.000        \\ \hline \hline
\end{tabular}
\end{center}
\caption{This table demonstrates
the errors and order accuracy of the proposed method 
for the $D_2$ disk for various number of zones $N=2^k$ from
$k=5$ to $10$ on polar coordinates.  
It shows that the order for the $D_2$ disk is about 1 in each norm. 
}
\label{tblPD2}
\end{table}

\begin{figure}
\begin{center}
\includegraphics[width=.32\textwidth]{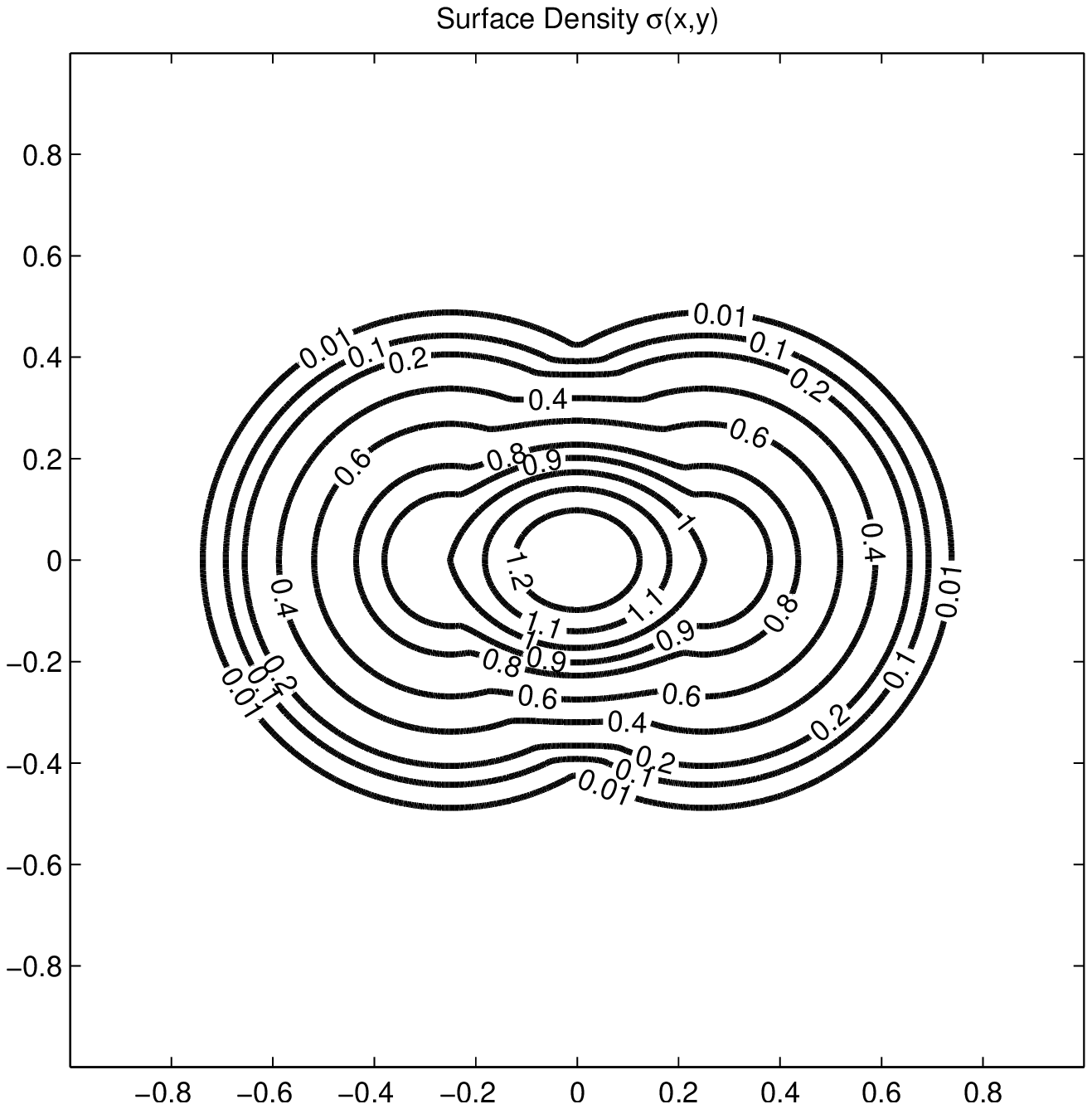}
\end{center}
\begin{center}
\includegraphics[width=.32\textwidth]{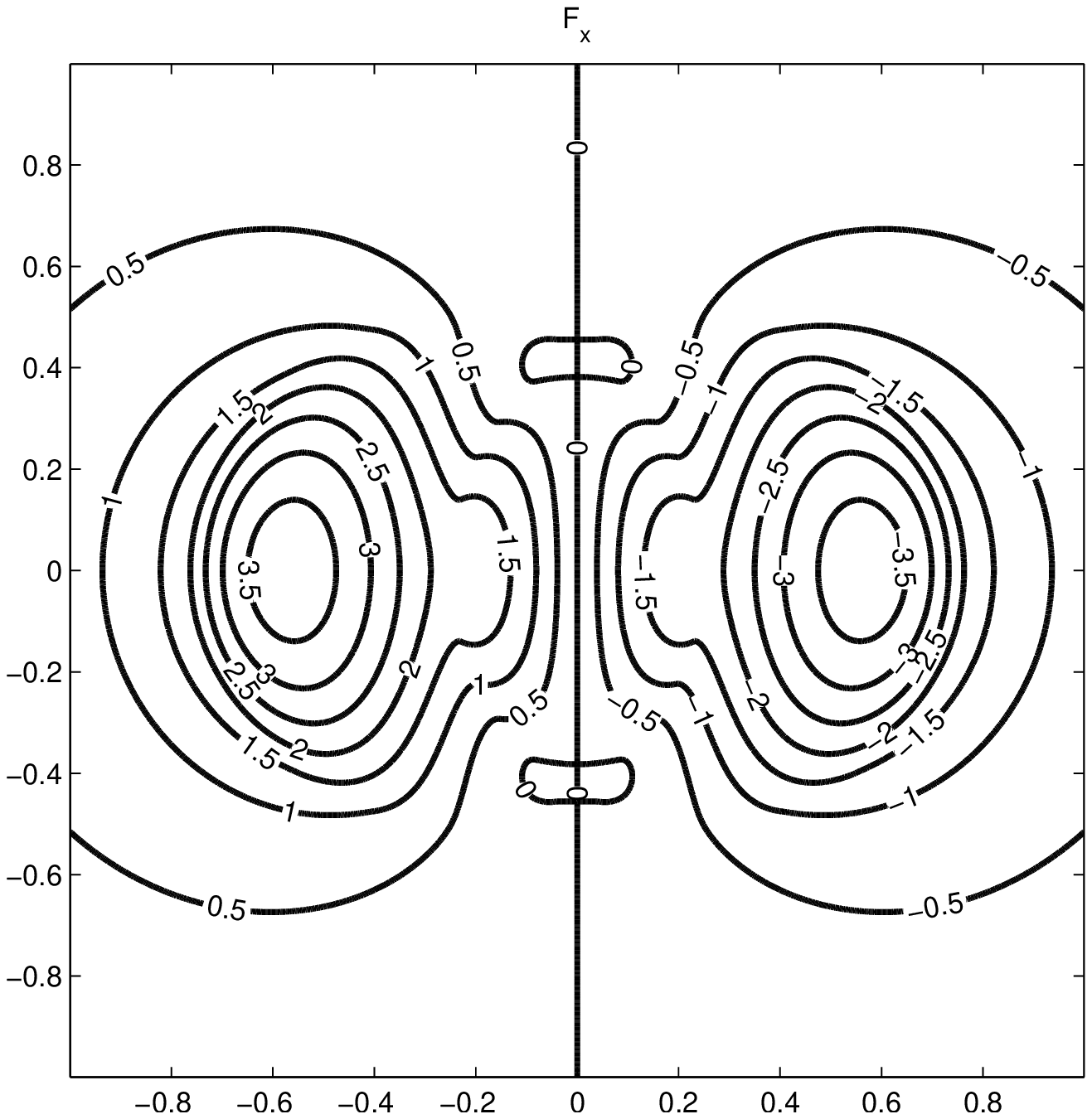}
\includegraphics[width=.32\textwidth]{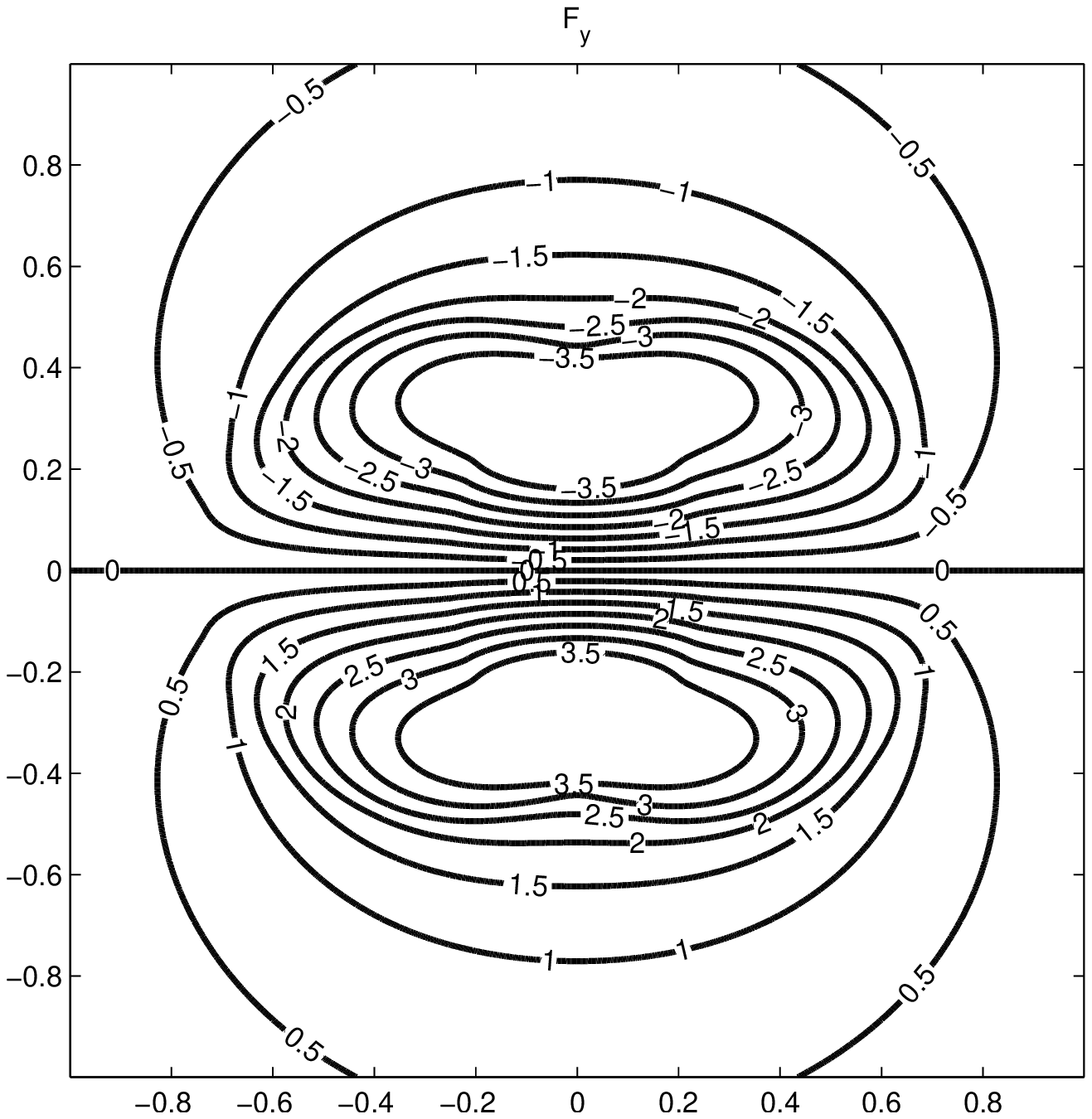}
\includegraphics[width=.32\textwidth]{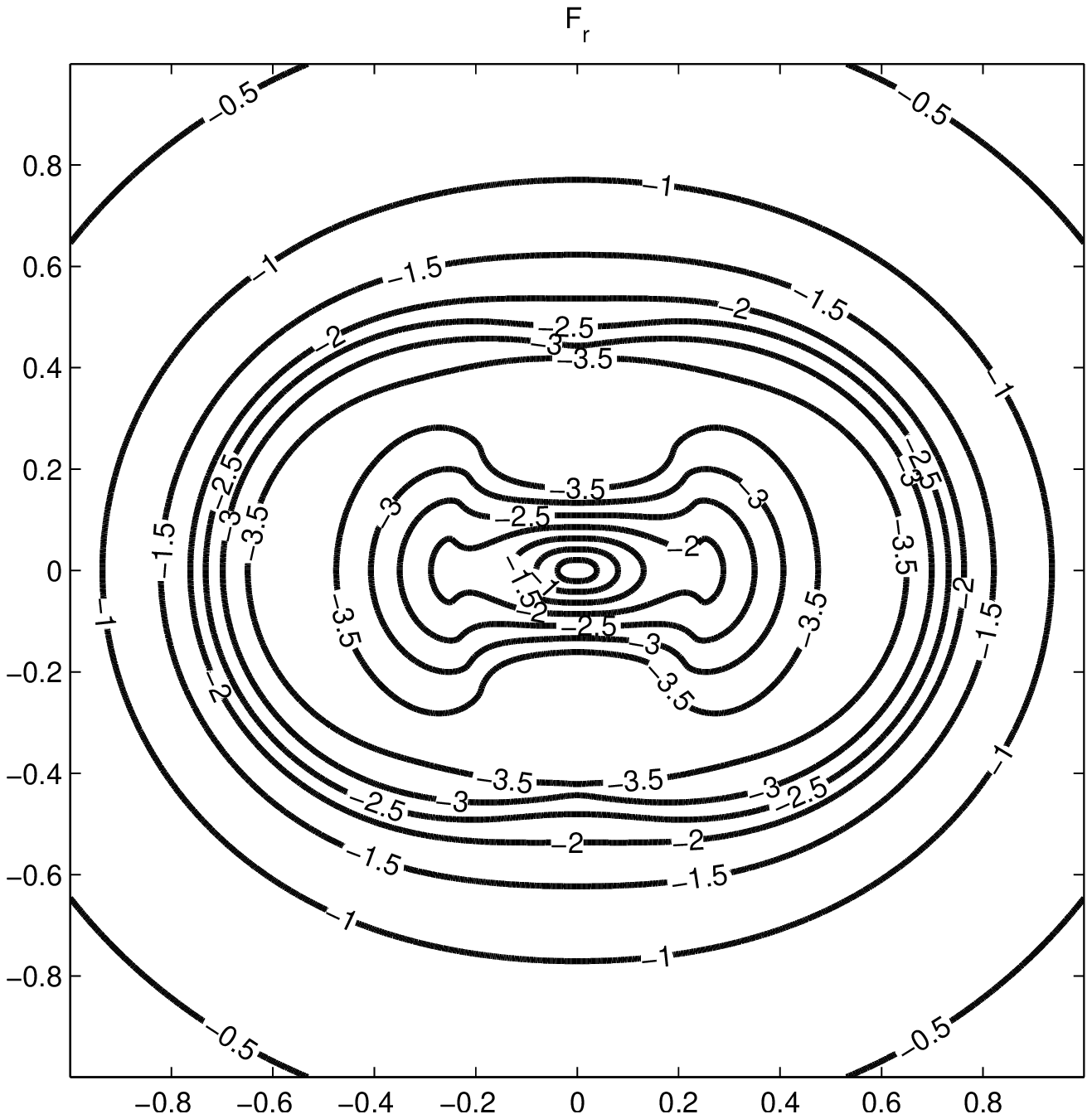}
\end{center}
\begin{center}
\includegraphics[width=.32\textwidth]{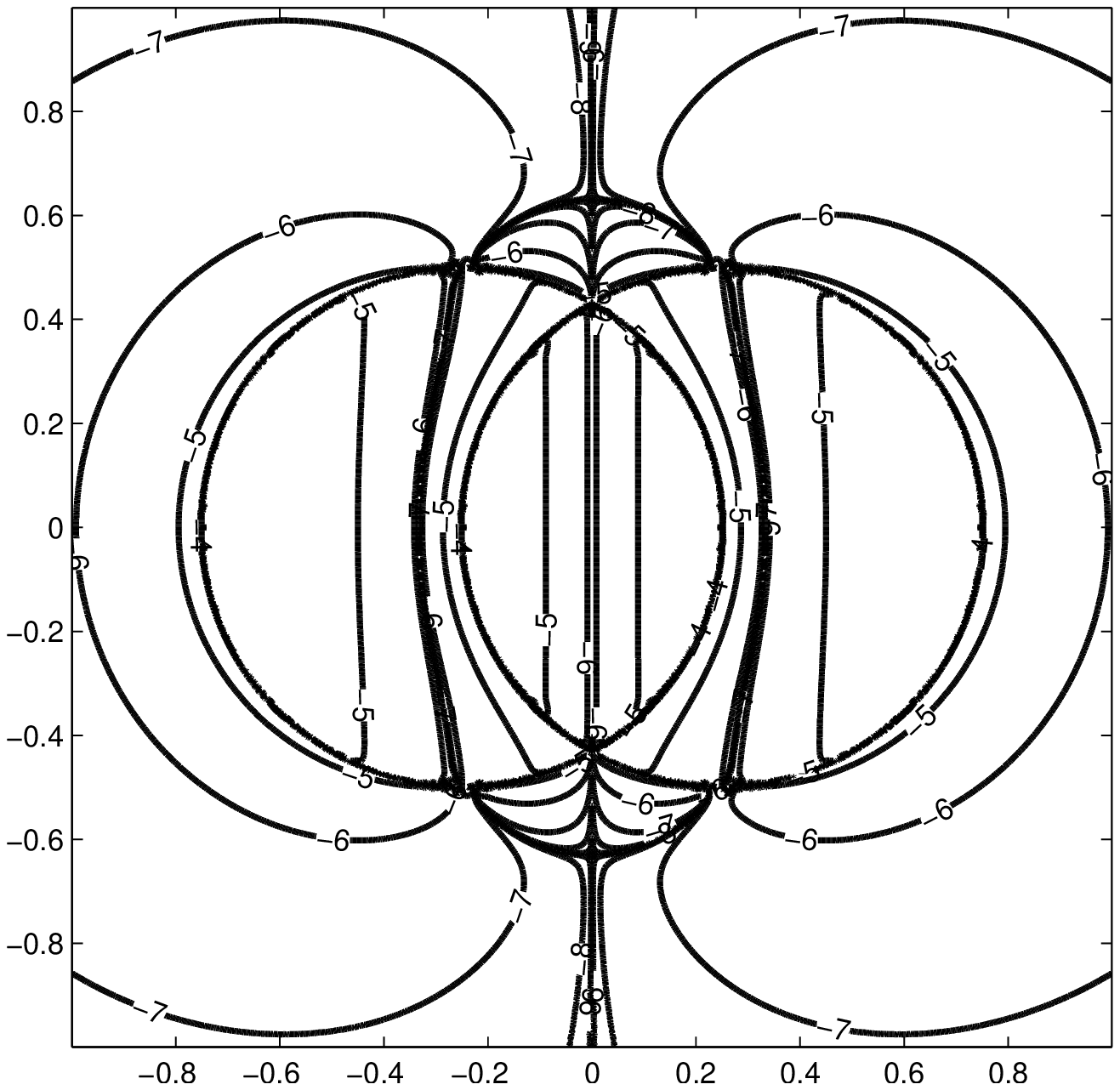}
\includegraphics[width=.32\textwidth]{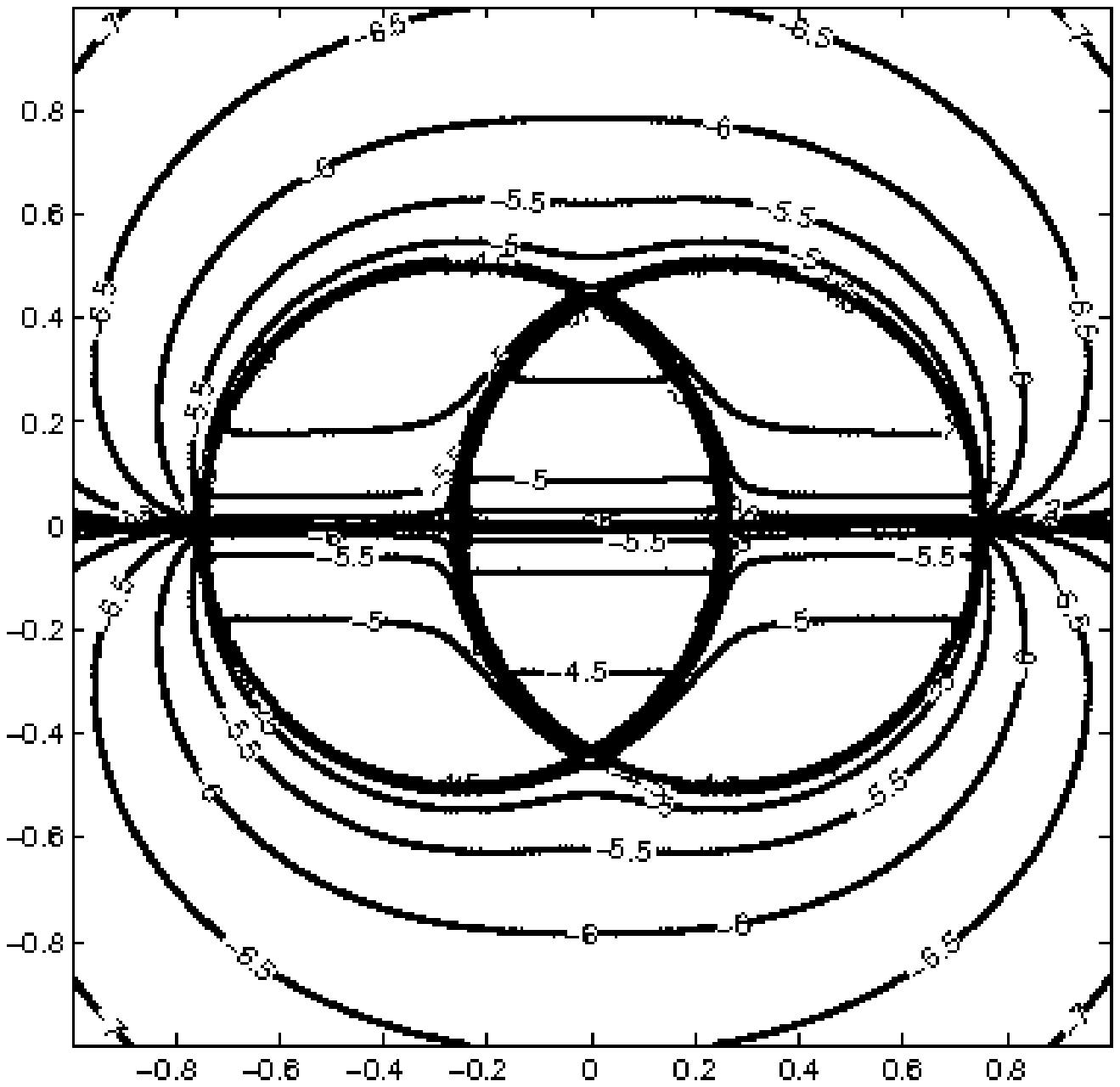}
\includegraphics[width=.32\textwidth]{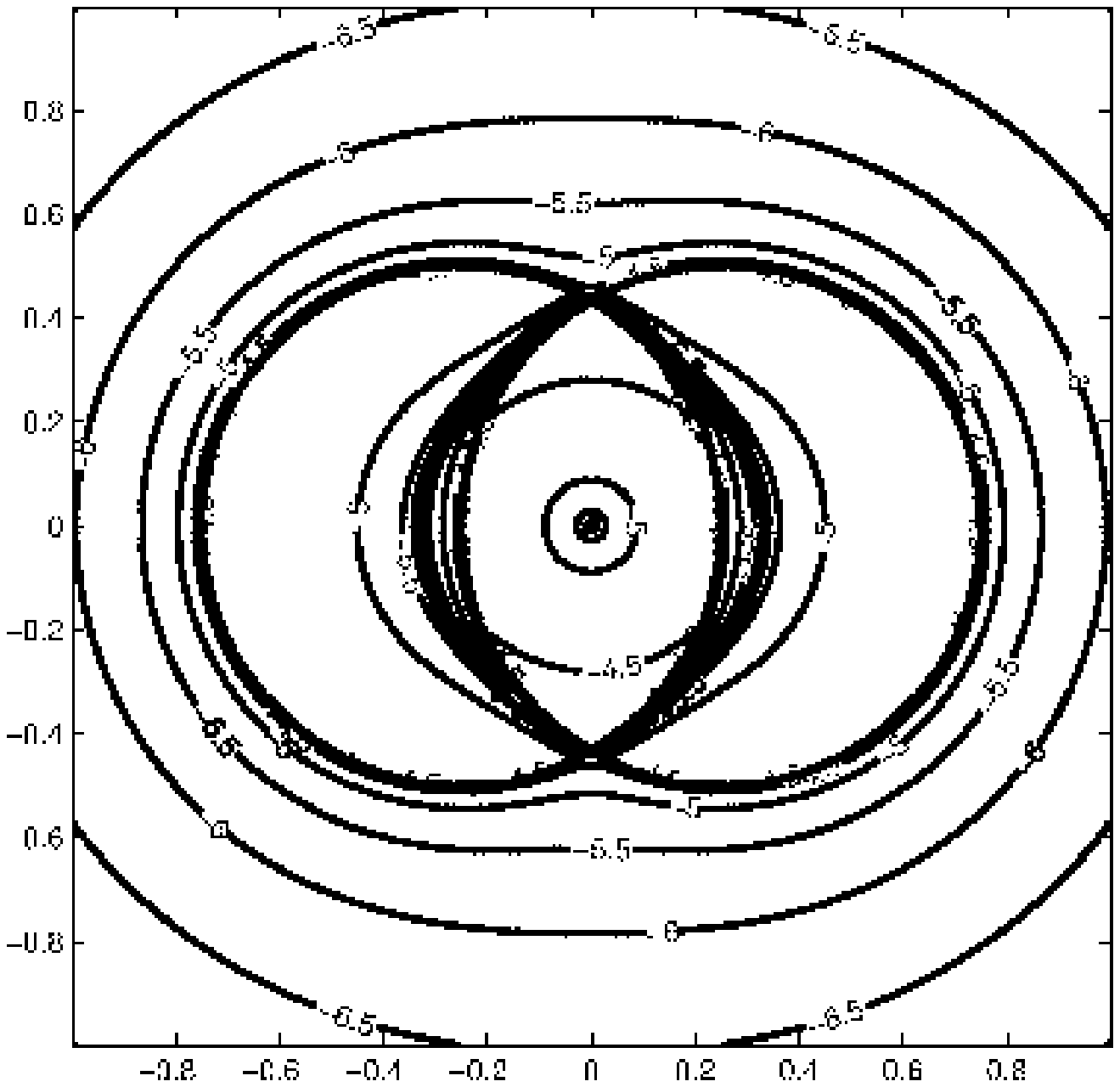}
\end{center}
\caption{The numerical solutions of a $D_{2,2}$ disk for $N=1024$,
The top contour plot is the surface density. 
The contour plots in the second row are the $x$-directional,
$y$-directional, and radial forces, respectively.
The corresponding errors between the numerical and analytic solutions in the third row.
The values in the  contour plots in the third row are 
the absolute errors in the common logarithmic scale. 
}
\label{DDN1024}
\end{figure}

\vskip 0.5cm
\noindent{\it Example 2}.
The disk $D_{2,2}$ of two superposed $D_2$ has the surface density 
$\Sigma_{D_{2,2}}= \Sigma_{D_2}(R_1;\alpha)+\Sigma_{D_2}(R_2;\alpha)$, where 
$R_1=\sqrt{(x-1/4)^2+y^2}$ and $R_2=\sqrt{(x+1/4)^2+y^2}$.
This example represents a non-symmetric distribution of the surface density of a disk. 
The results are shown in Table \ref{tblD2D2} and Figure \ref{DDN1024}. 
This result is similarly to Example 1.
The factors $O^\infty_x$ of errors in Table \ref{tblD2D2}
are non-monotonic as the numerical resolution, $N_k$, increases.
This may be due to the distribution of the surface density on grid cells, 
the centers of which can shift with varying numerical resolution.
However, the total variation and energy shows the convergence and
the order of accuracy is about 1.8 and 1.9 respectively.

\begin{table}
\begin{center}
\begin{tabular}{|c|c|c|c||c|c|c||c|c|c|}  \hline
$N$ & $E^1_x$   & $E^2_x$   & $E^\infty_x$  & $E^1_y$ & $E^2_y$ & $E^\infty_y$ & $E^1_R$   &  $E^2_R$  & $L^\infty_R$ \\ \hline 
32	&1.56E-2	&1.29E-2	&2.19E-2	    & 2.09E-2 & 1.71E-2 & 3.57E-2      &2.56E-2    &1.95E-2	&3.55E-2 \\ \hline
64	&4.29E-3	&3.83E-3	&7.75E-3	    & 5.38E-3 & 4.57E-3 & 1.07E-2      &6.89E-3	   &5.53E-3	&1.16E-2 \\ \hline
128	&1.23E-3	&1.18E-3	&5.41E-3	    & 1.50E-3 & 1.35E-3 & 5.72E-3      &1.96E-3	   &1.64E-3	&5.81E-3 \\ \hline
256	&3.17E-4	&3.12E-4	&1.94E-3	    & 3.83E-4 & 3.54E-4 & 1.96E-3      &5.06E-4	   &4.34E-4	&2.01E-3 \\ \hline
512	&8.32E-5	&9.00E-5	&9.49E-4	    & 9.99E-5 & 9.89E-5 & 9.53E-4      &1.33E-4	   &1.23E-4	&9.64E-4 \\ \hline
1024&2.12E-5	&2.41E-5	&3.28E-4	    & 2.54E-5 & 2.62E-5 & 3.29E-4      &3.39E-5	   &3.29E-5	&3.38E-4 \\ \hline \hline
$N$ &   $O^1_x$   & $O^2_x$   & $O^\infty_x$  & $O^1_y$& $O^2_y$& $O^\infty_y$& $O^1_R$   &  $O^2_R$  & $O^\infty_R$ \\ \hline
32/64	&1.86	&1.75	&1.50                 & 1.96   &  1.87  &  1.74         &1.89	&1.82	&1.62 \\ \hline
64/128	&1.80	&1.71	&0.52	              & 1.85   &  1.79  &  0.90         &1.81	&1.75	&1.00 \\ \hline
128/256	&1.96	&1.91	&1.48	              & 1.97   &  1.93  &  1.55         &1.95	&1.92	&1.53 \\ \hline
256/512	&1.93	&1.79	&1.03	              & 1.94   &  1.84  &  1.04         &1.93	&1.81	&1.06 \\ \hline
512/1024&1.97	&1.90	&1.53	              & 1.98   &  1.91  &  1.53         &1.97	&1.90	&1.51 \\ \hline
\end{tabular}
\end{center}
\caption{This table demonstrates
the errors and order of accuracy of the proposed method 
for the $D_{2,2}$ disk for various number of zones $N=2^k$ from
$k=5$ to $10$.  
It shows that the order for the $D_{2,2}$ disk is about $1.8$ or
$1.9$ order in $L^1$ and $L^2$ norm. 
}
\label{tblD2D2}
\end{table}

\begin{figure}
\begin{center}
\includegraphics[width=.42\textwidth]{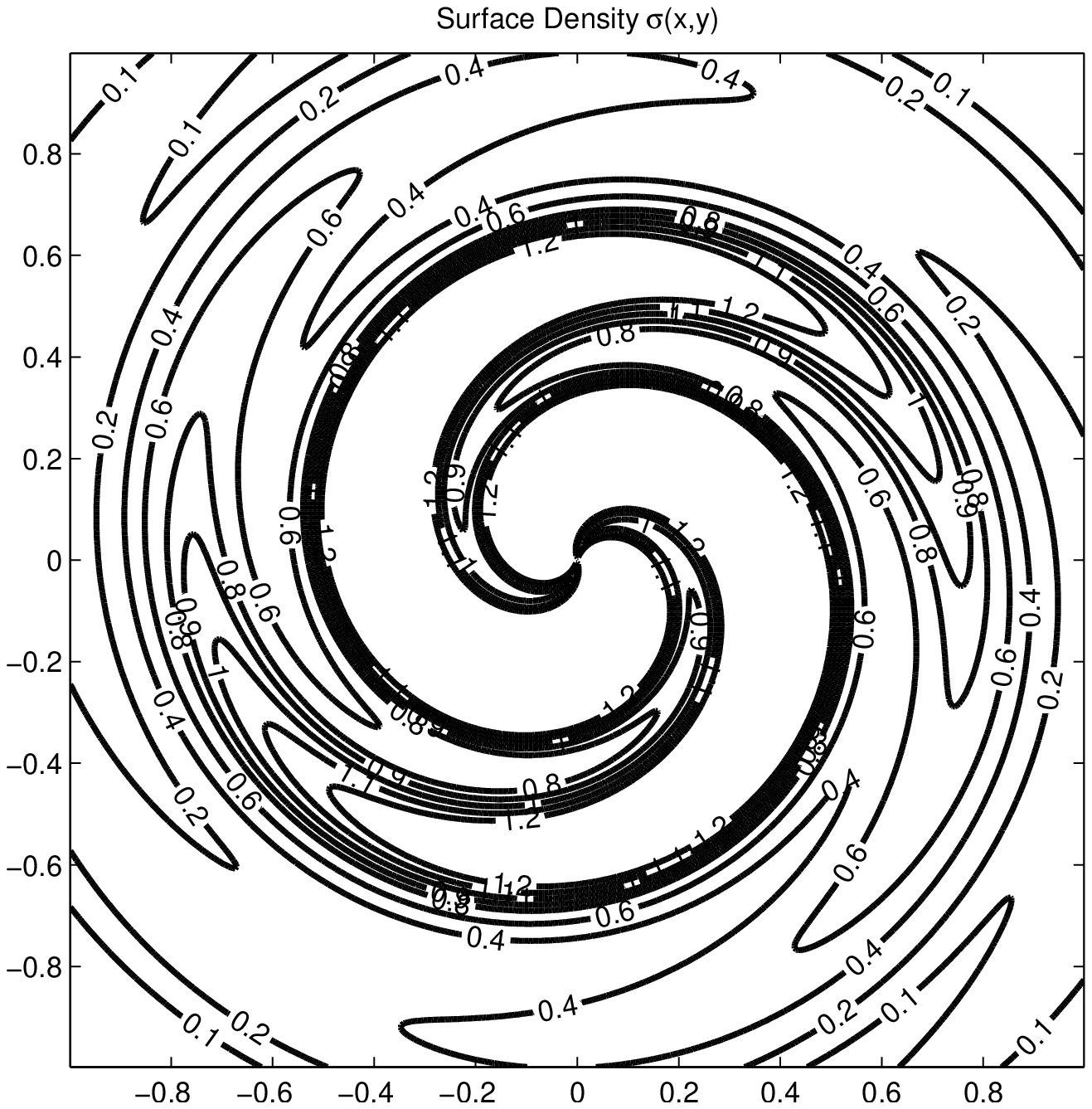}
\includegraphics[width=.42\textwidth]{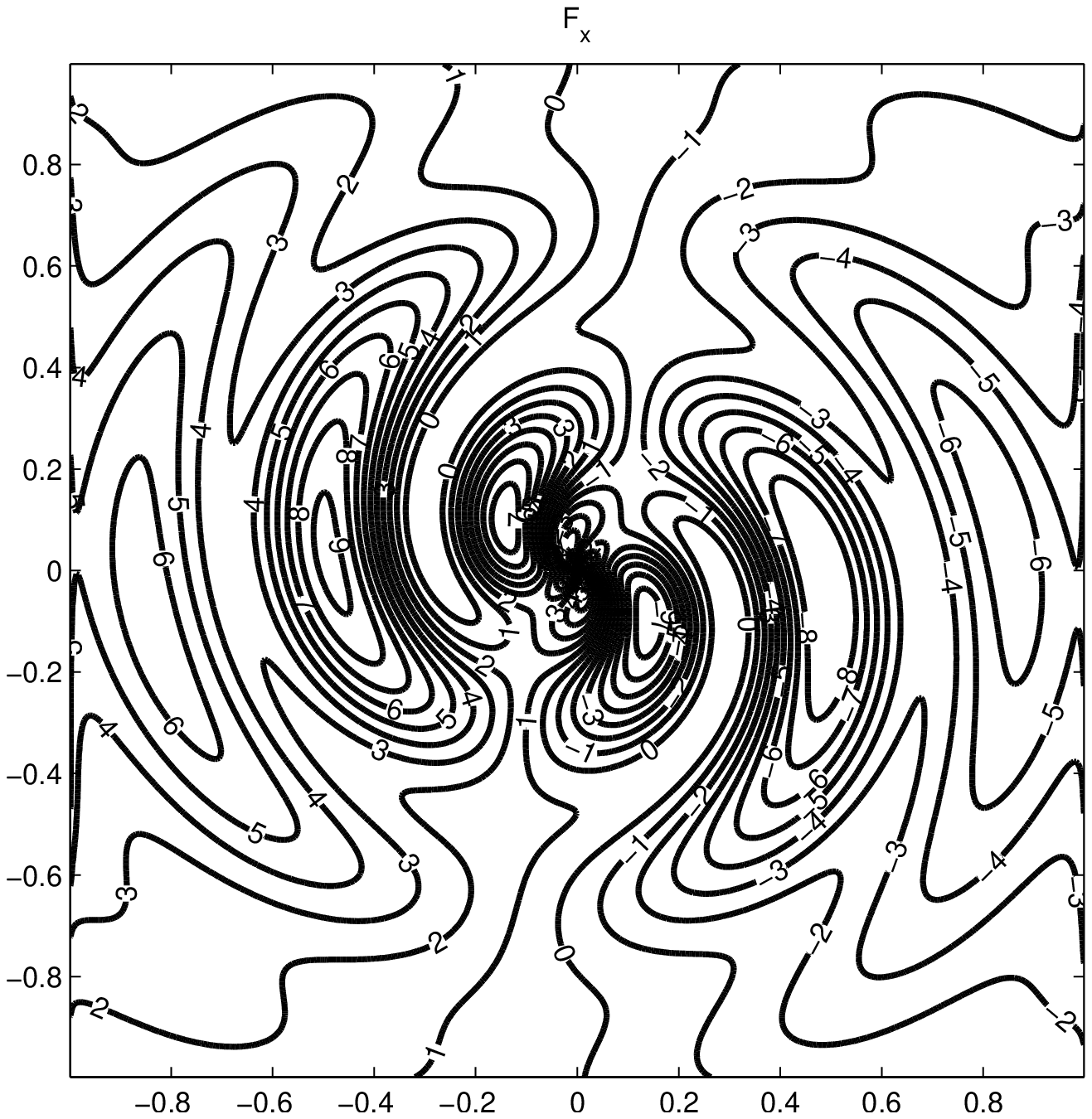}
\end{center}
\begin{center}
\includegraphics[width=.42\textwidth]{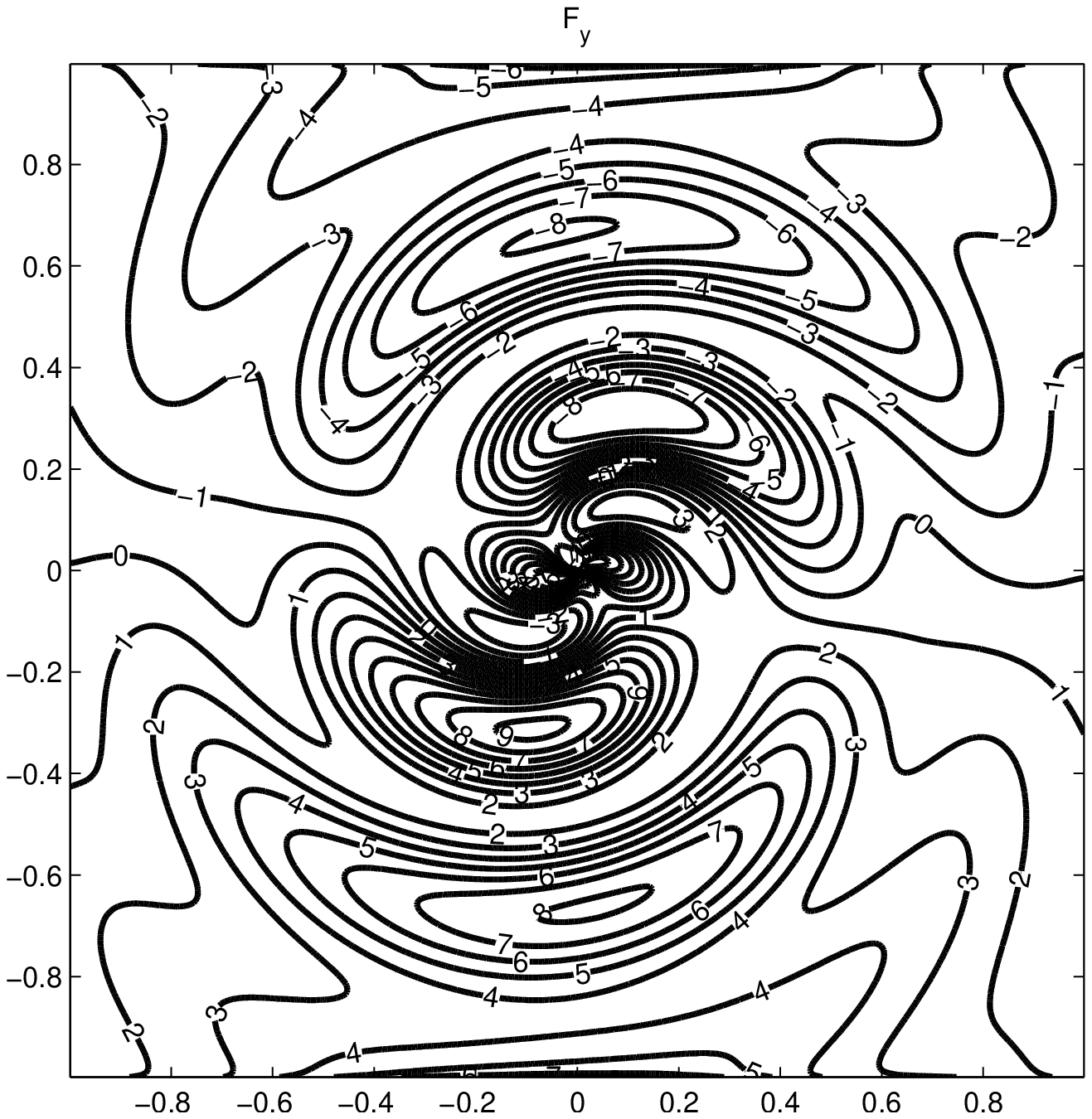}
\includegraphics[width=.42\textwidth]{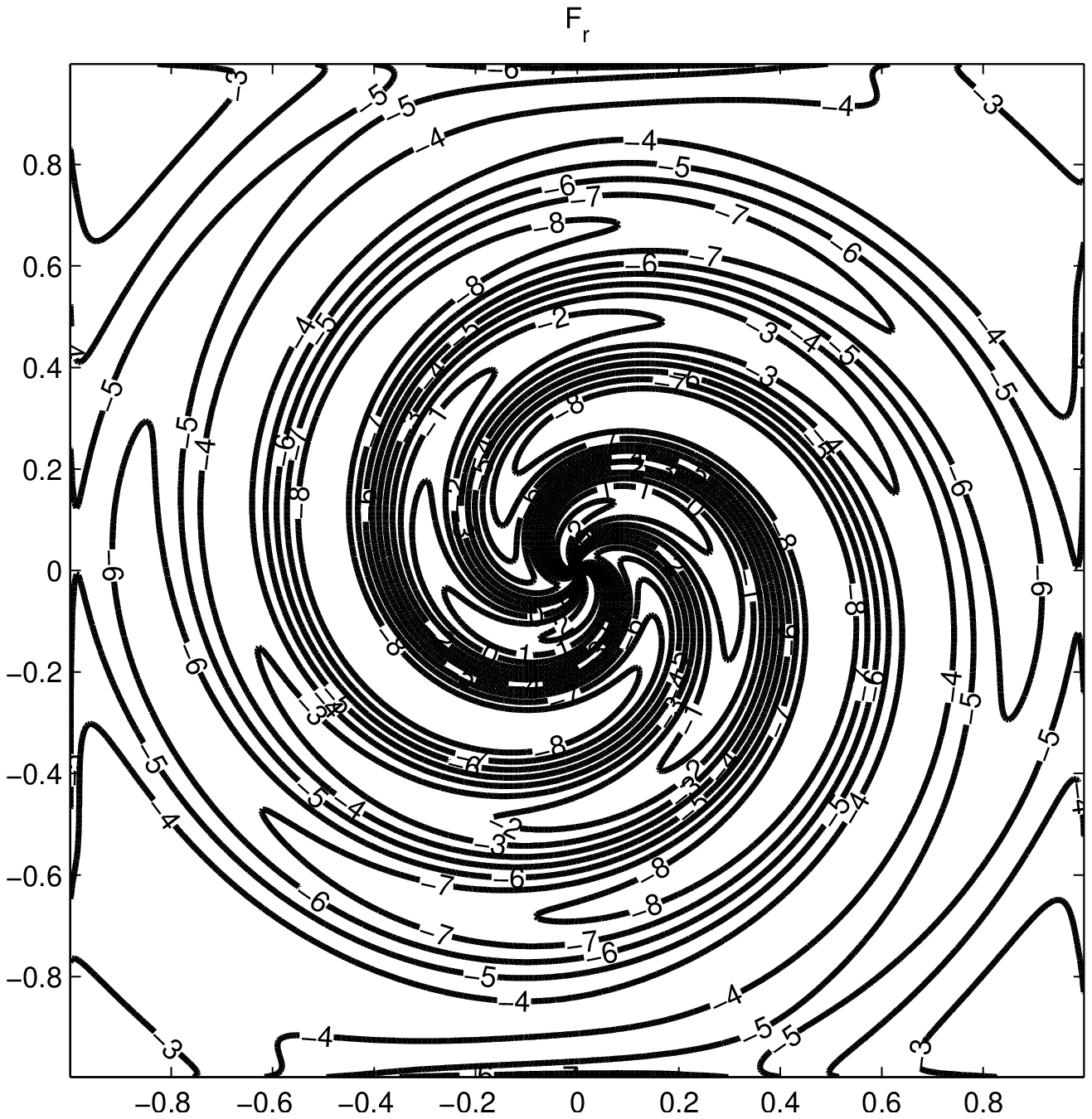}
\end{center}
\caption{The numerical solutions of a logarithmic spiral disk for $N=512$
to investigate the self-gravitational force calculation.
The contour plots illustrate the surface density (upper left), 
$x$-force (upper right), $y$-force (low left), and radial force (lower right).}
\label{EX3CPSD}
\end{figure}

\vskip 0.5cm
\noindent{\it Example 3}. As another example of a non-axisymmetric potential, we consider a logarithmic 
spiral disk.  Since an analytic pair for the surface density and potential are not known, we assume a surface
density profile of the form
\begin{eqnarray*}
\Sigma_{LS}(r,\theta)=e^{-2r^2}(2+\cos(2\theta+16 r)).
\end{eqnarray*}
To investigate the order of accuracy, the solution at the finest mesh size is regarded as 
the true solution. 
For various coarser resolutions, the values at some specific position
are taken to be the average of the four closest to the position.
The results are shown for the method based on Cartesian coordinates in 
Table \ref{tblCS3} and Figure \ref{EX3CPSD}. 
It can be seen that the order of accuracy is about $1.5$ for the $L^1$ norm 
and about $1$ for the $L^2$ norm. The $L^\infty$ norm is only convergent.

\begin{table}
\begin{center}
\begin{tabular}{|c|c|c|c||c|c|c||c|c|c|}  \hline
$N$ & $E^1_x$   & $E^2_x$   & $E^\infty_x$  & $E^1_y$ & $E^2_y$ & $E^\infty_y$ & $E^1_R$   &  $E^2_R$  & $L^\infty_R$ \\ \hline 
32	&3.40E-1	&2.97E-1	&1.40E-0	    & 3.38E-1 & 2.98E-1 & 1.39E-0      &4.64E-1    &3.03E-1	&4.27E-1\\ \hline
64	&1.21E-1	&1.70E-1	&1.83E-0	    & 1.23E-1 & 1.72E-1 & 1.83E-0      &1.36E-1	   &9.29E-2	&2.07E-1\\ \hline
128	&4.71E-2	&9.28E-2	&1.92E-0	    & 4.83E-2 & 9.51E-2 & 1.92E-0      &3.97E-2	   &3.93E-2	&2.02E-1\\ \hline
256	&1.87E-2	&4.50E-2	&1.71E-0	    & 1.93E-2 & 4.68E-2 & 1.71E-0      &1.17E-2	   &2.11E-2	&1.75E-1\\ \hline
512	&6.05E-3	&1.67E-2	&1.16E-0	    & 6.30E-3 & 1.78E-2 & 1.16E-0      &1.00E-3	   &9.53E-3	&1.54E-1 \\ \hline \hline
$N$ &   $O^1_x$   & $O^2_x$   & $O^\infty_x$  & $O^1_y$& $O^2_y$& $O^\infty_y$& $O^1_R$   &  $O^2_R$  & $O^\infty_R$ \\ \hline
32/64	&1.49	&0.81	&-0.39                & 1.46   &  0.79  &  -0.40        &1.77	&1.70	&1.05 \\ \hline
64/128	&1.36	&0.87	&-0.07	              & 1.34   &  0.85  &  -0.07        &1.77	&1.24	&0.03 \\ \hline
128/256	&1.34	&1.05	&0.17	              & 1.32   &  1.02  &   0.17        &1.76	&0.90	&0.21 \\ \hline
256/512	&1.63	&1.43	&0.56	              & 1.62   &  1.62  &   0.56        &1.96	&1.14	&0.18 \\ \hline
\end{tabular}
\end{center}
\caption{This table demonstrates
the errors and order of accuracy of the proposed method 
for the spiral disk for various number of zones $N=2^k$ from
$k=5$ to $9$.   
It shows that the order for the spiral disk is about 1.5 or 1.0 in the $L^1$ and $L^2$ norms, respectively.
}
\label{tblCS3}
\end{table}

\subsection{A comparison study}
The goal of this paper is to calculate the self-gravitational forces with as few restrictions 
as possible.  The most straight forward approach is to solve for the potential via (\ref{DPG}) and 
obtain the self-gravitational forces by taking its derivatives. 
If one uses the finite difference or finite element method on (\ref{DPG}), the discretization is 
\begin{eqnarray*}
\frac{-\Phi_{i+1,j  ,k  }+2\Phi_{i,j,k}-\Phi_{i-1,j  ,k  }}{(\Delta x)^2}+
\frac{-\Phi_{i  ,j+1,k  }+2\Phi_{i,j,k}-\Phi_{i  ,j-1,k  }}{(\Delta y)^2}+
\frac{-\Phi_{i  ,j  ,k+1}+2\Phi_{i,j,k}-\Phi_{i  ,j  ,k-1}}{(\Delta z)^2}
=-f_{i,j,k}
\end{eqnarray*}
where $\Phi_{i,j,k}=\Phi(x_i,y_j,z_k)$ and $f_{i,j,k}=f(x_i,y_j,z_k)$
based on the uniform mesh grids $(x_i,y_j,z_k)$. Here, 
$f_{i,j,k}=0$ for $k\not=0$.
For such an approach, artificial boundary conditions should be imposed and a fully 3-dimensional 
calculation must be undertaken. We point out that the (\ref{DPG}) can not be reduced to the two 
dimensional numerical partial differential problem, i.e.,
 \begin{eqnarray*}
\frac{-\Phi_{i+1,j,  0}+2\Phi_{i,j,0}-\Phi_{i-1,j  ,0}}{(\Delta x)^2}+
\frac{-\Phi_{i  ,j+1,0}+2\Phi_{i,j,0}-\Phi_{i  ,j-1,0}}{(\Delta y)^2}
=-f_{i,j,0}.
\end{eqnarray*}
Any numerical solution of the partial differential problem will involve $O(N^3)$ unknowns. 
It follows that the linear complexity of such an approach, viz. multigrid method, is at least 
$O(N^3)$.  For an infinitesimally thin gaseous disk problem, 
this approach does not appear to be suitable.  

\begin{figure}
\begin{center}
\includegraphics[width=.42\textwidth]{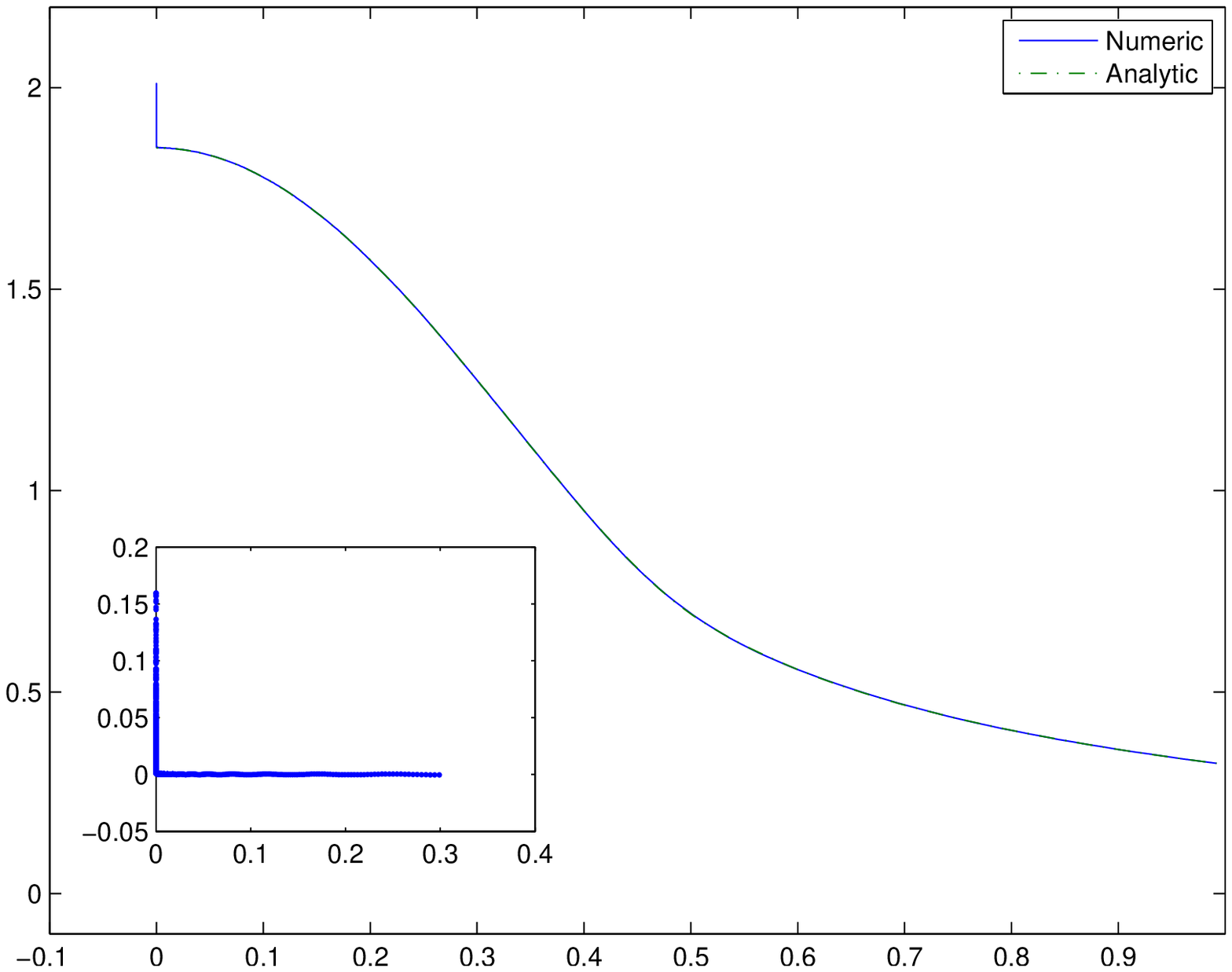}
\includegraphics[width=.42\textwidth]{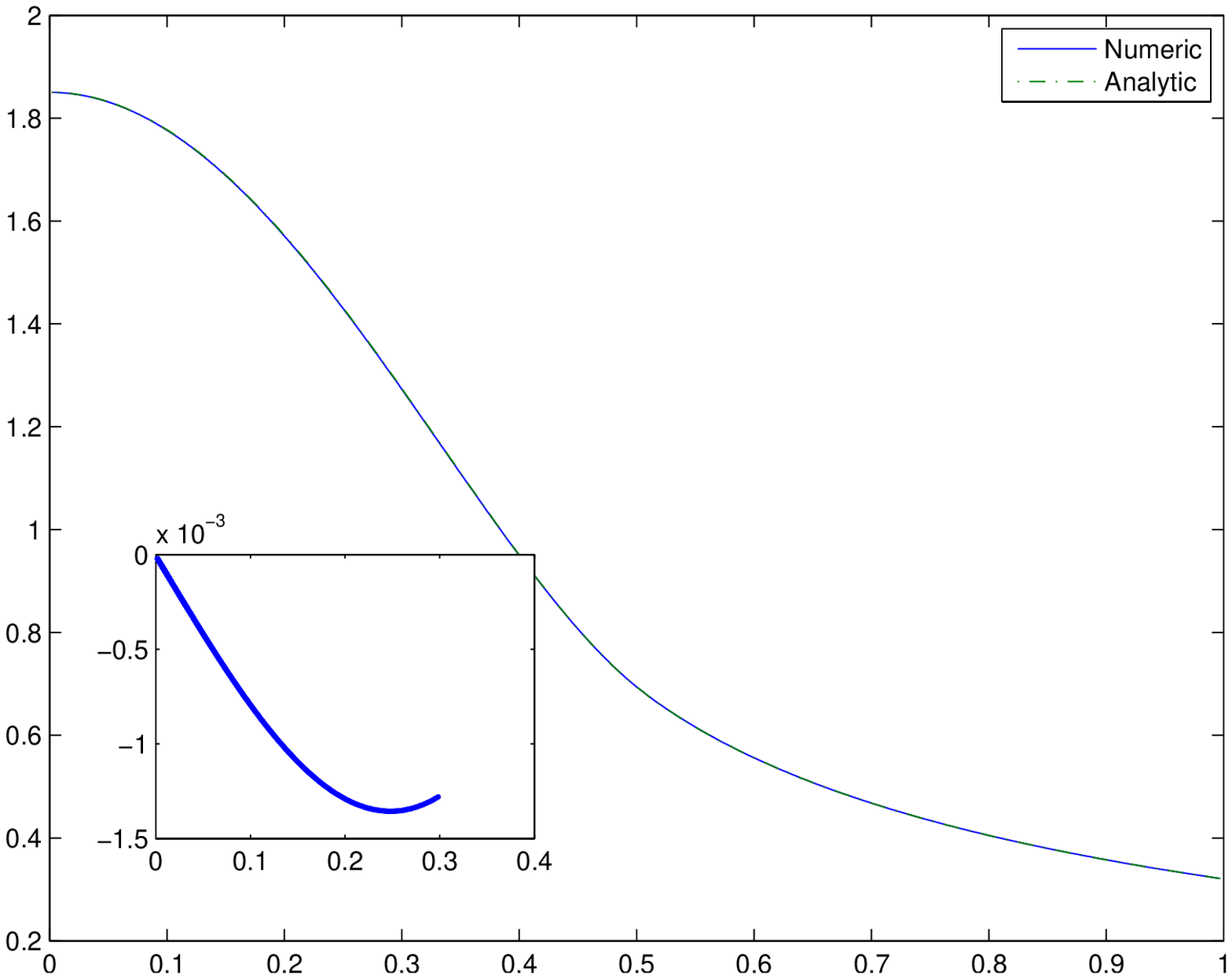}
\end{center}
\caption{The variation of the potential with respect to radius using Kalnajs' method (left) and
the proposed method (right). The residuals are shown in the small window in each panel and 
show that the Kalnajs' method have significant errors near the origin, which are eliminated in 
the proposed method. }
\label{Kalnajs1D}
\end{figure}

Alternatively, one can solve the reduced equation given by
\begin{eqnarray*}
\Phi(x,y,0)=-G\int\!\!\!\int
\frac{\sigma(\bar x,\bar y)}{\sqrt{(\bar x-x)^2+(\bar y-y)^2}}
d\bar x d\bar y
\end{eqnarray*}
or 
\begin{eqnarray*}
\Phi(r,\theta,0)=-G\int\!\!\!\int
\frac{\sigma(\bar r,\bar \theta)}{\sqrt{{\bar r}^2+r^2-2\bar r r \cos(\bar\theta-\theta)}}
\bar r d\bar r d\bar \theta.
\end{eqnarray*}
In this case, one can consider using bases functions on a two dimensional space as in a 
spectral method.  Unfortunately, this approach requires a treatment for the boundary conditions. 
A possible way to deal with this issue is to impose periodic boundary conditions. 
However, it is not realistic for a gravitational force calculation because gravity is a long range 
force and not periodic. 
As an alternative, a method without the periodic assumption has been proposed for 
polar coordinates~\cite{Kalnajs1971}. The approach in \cite{Kalnajs1971}
transforms the polar coordinate $(r,\theta)$ into the coordinate $(u,\theta)$
by setting $r=e^u$ or $u=\ln(r)$. The potential-density pair 
in term of the reduced surface density and reduced potential is given in \cite{Kalnajs1971}, 
and it is
\begin{eqnarray*}
e^{3u/2}\sigma(e^u,\theta)=\frac{1}{4\pi^2}\sum_m \int^\infty_{-\infty}A_m(\alpha)e^{i(m\theta+\alpha u)}d\alpha
\end{eqnarray*}
and
\begin{eqnarray}
\label{uPotentials}
e^{u/2}\Phi(e^u,\theta) = -\frac{1}{2\pi}G\sum_m \int^\infty_{-\infty}K(\alpha,m)A_m(\alpha)
\exp[i(m\theta+\alpha u)]d\alpha,
\end{eqnarray} 
where $K$ is real and positive and is defined as 
\begin{eqnarray*}
K(\alpha,m)\equiv \frac{1}{2}
\frac{\Gamma[(m+1/2+i\alpha)/2]\Gamma[(m+1/2-i\alpha)/2]}{\Gamma[(m+3/2+i\alpha)/2]\Gamma[(m+3/2-i\alpha)/2]}.
\end{eqnarray*}
We regard this method as one of the spectral methods because Fourier series $e^{-im\theta}$
and Fourier integral $e^{-i\alpha u}$ are used. 
To apply this method to the $D_2$ disk using the polar coordinates, we transform the 
bounded unit disk $D(0,1)=[0,1]\times [0,2\pi]$ to 
the unbounded domain $U=(-\infty,0]\times [0,2\pi]$. 
In this special case, we only need to compute $m=0$ and truncate  
\begin{eqnarray}
\label{Amalpha}
A_0(\alpha)=\int^0_{-\infty} e^{3u/2}\sigma(e^u)e^{-i\alpha u}du
           \approx \int^0_{u_{\min}} e^{3u/2}\sigma(e^u)e^{-i\alpha u}du,
\end{eqnarray}
where the value $u_{\min}$ is to approximate $-\infty$. 
The truncation produces a hole in the unit disk and can introduce significant errors at the origin.
Given a positive integer $N$ and 
base on the discretization for the radial region in the previous subsection, 
to calculate (\ref{Amalpha}) and (\ref{uPotentials}) by the trapzoidal rule.
The variation of the potential with respect to radius is illustrated in 
Figure~\ref{Kalnajs1D}.  The profile on the left panel shows that the numerical 
and analytic solutions for the Kalnajs' method agree well except close to the origin
for $N=1024$.  The small window embedded within the panel zooms in on the residuals 
between numerical and analytic solution on the interval $[0,0.3]$. It is seen that the 
truncated portion contributes to significant errors near the origin.  In contrast, the 
application of our proposed method to the calculation of potentials leads to the results 
shown in the right panel of Figure~\ref{Kalnajs1D}. Although the singular integration 
still remains due to the unbounded domain, our proposed method on either Cartesian and 
polar coordinates is preferable since a hole near the origin is not introduced.

Finally, a third approach is to directly calculate the integrals and obtain the potential. 
For any given mesh grid, the total amount of complexity is $O(N^4)$ based on the number 
$O(N^2)$ of mesh zones.  If we restrict ourselves to a uniform grid and use the FFT technique,
the complexity can be reduced from $O(N^4)$ to $O(N^2)$.  
In other words, a fast algorithm of linear complexity is obtained.
It is common to start with 
\begin{eqnarray*}
\Phi(x,y,0)
&=&-G\int\!\!\!\int 
{\cal K}(\bar x-x, \bar y-y,0)\sigma(\bar x,\bar y) d\bar x d\bar y\\
&=&-G\sum^N_{i=1}\sum^N_{j=1}
\int\!\!\!\int_{D_{i,j}}
{\cal K}(\bar x-x, \bar y-y,0)\sigma(\bar x,\bar y) d\bar x d\bar y.\\
\end{eqnarray*}
and to introduce a softening parameter $\epsilon$ to approximate 
\begin{eqnarray*}
\int\!\!\!\int_{D_{i,j}}{\cal K}(\bar x-x,\bar y-y)\sigma(\bar x,\bar y)
\approx -\frac{G}{\sqrt{\epsilon^2+(x_{i'}-x_i)^2+(y_{j'}-y_j)^2}}
\int\!\!\!\int_{D_{i,j}}\sigma(\bar x,\bar y)d\bar x d\bar y.
\end{eqnarray*} 
Since the goal is to calculate the forces,  
the order of accuracy is reduced when taking 
the numerical differentiation on the numerical solution of potentials.
For polar coordinates~\cite{Binney}, the value of ${\cal K}$ is approximated by
\begin{eqnarray*}
{\cal K}_{i'-i,j'-j}
:= -\frac{G}{\sqrt{2(\cosh(u_{i'}-u_i)-\cos(\theta_{j'}-\theta_j))}},
\end{eqnarray*}
where $u_{i'}=\ln(x_{i'})$ and $u_i=\ln(x_i)$.
Note that when $(i',j')=(i,j)$, ${\cal K}$ is undefined.  
An approach to avoid the singularity 
problem can be found in~\cite{Binney}.  
On the other hand, the proposed method avoids the 
singularity problem by directly evaluating the forces, hence, raising the order of accuracy. 
For Cartesian coordinates, we choose the softening parameters as 
the mesh size $\epsilon=\Delta x$. The errors for the disks $D_2$
and $D_{2,2}$ are shown in Table~\ref{tblSPCD2}
and Table~\ref{tblSPCD2D2}, respectively.
It reveals that the accuracy when using 
the softening parameter approach for the $D_2$ and $D_{2,2}$ disks
is of first order in the $L^1$ and $L^2$ norms. 
For the $L^\infty$ norm, the order of accuracy for the $D_2$ disk
is about $1$. For the $D_{2,2}$ disk, this method loses accuracy.   
In comparison  with our proposed method for Example 1 and Example 2, 
our methods are more accurate and the order of accuracy is verified. 

\begin{table}
\begin{center}
\begin{tabular}{|c|c|c|c||c|c|c|c|}  \hline
$N$ & $E^1_x$   &  $E^2_x$  & $L^\infty_x$ & $N_{k-1}/N_k$ &  $O^1_x$ & $O^2_x$ & $O^\infty_x$ \\ \hline 
32	&4.283E-1   &5.116E-1	&9.981E-1      &               &          &         &              \\ \hline
64	&2.223E-1	&2.768E-1	&5.415E-1      & 32/64         & 0.9461   & 0.8862  &  0.9377      \\ \hline
128	&1.133E-1	&1.442E-1	&2.827E-1      & 64/128        & 0.9724   & 0.9408  &  0.9377      \\ \hline
256	&5.721E-2	&7.364E-2	&1.440E-1      & 128/256       & 0.9858   & 0.9695  &  0.9732      \\ \hline
512	&2.874E-2	&3.722E-2	&7.282E-2      & 256/512       & 0.9932   & 0.9844  &  0.9837      \\ \hline
1024&1.440E-2   &1.871E-2   &3.659E-2      & 512/1024      & 0.9970   & 0.9923  &  0.9929      \\ \hline \hline
\end{tabular}
\end{center}
\caption{This table demonstrates
the errors and order accuracy of the softening parameter  method 
for the $D_2$ disk for various number of zones $N=2^k$ from
$k=5$ to $10$ in Cartesian coordinates.  
It shows that the accuracy for the $D_2$ disk is about first order.}
\label{tblSPCD2}
\end{table}

\begin{table}
\begin{center}
\begin{tabular}{|c|c|c|c||c|c|c||c|c|c|}  \hline
$N$ & $E^1_x$   & $E^2_x$   & $E^\infty_x$  & $E^1_y$ & $E^2_y$ & $E^\infty_y$ & $E^1_R$   &  $E^2_R$  & $L^\infty_R$ \\ \hline 
32	&5.95E-1	&5.61E-1	&1.00E-0	    & 9.13E-1 & 8.31E-2 & 1.46E-0      &1.16E-0    &9.32E-1	&1.45E-0 \\ \hline
64	&3.10E-1	&3.10E-1	&5.42E-1	    & 4.73E-1 & 4.49E-3 & 8.04E-1      &5.97E-1	   &5.06E-1	&8.04E-1 \\ \hline
128	&1.59E-1	&1.69E-1	&4.17E-1	    & 2.41E-1 & 2.36E-3 & 4.22E-1      &3.03E-1	   &2.69E-1	&4.21E-1 \\ \hline
256	&8.04E-2	&9.29E-2	&4.17E-1	    & 1.22E-1 & 1.24E-4 & 3.02E-1      &1.53E-1	   &1.44E-1	&4.17E-1 \\ \hline
512	&4.05E-2	&5.31E-2	&4.17E-1	    & 6.10E-2 & 6.57E-5 & 3.03E-1      &7.68E-2	   &7.85E-2	&4.17E-1 \\ \hline
1024&2.03E-2	&3.19E-2	&4.17E-1	    & 3.06E-2 & 3.61E-5 & 3.03E-1      &3.85E-2	   &4.49E-2	&4.17E-1 \\ \hline \hline
$N$ &   $O^1_x$   & $O^2_x$   & $O^\infty_x$  & $O^1_y$& $O^2_y$& $O^\infty_y$& $O^1_R$   &  $O^2_R$  & $O^\infty_R$ \\ \hline
32/64	&0.94	&0.86	&0.88                 & 0.95   &  0.89  &  0.86         &0.95	&0.88	&0.85 \\ \hline
64/128	&0.97	&0.88	&0.38	              & 0.97   &  0.93  &  0.93         &0.98	&0.91	&0.93 \\ \hline
128/256	&0.98	&0.86	&0.00	              & 0.99   &  0.93  &  0.47         &0.99	&0.90	&0.02 \\ \hline
256/512	&0.99	&0.80	&0.00	              & 0.99   &  0.91  &  0.00         &0.99	&0.87	&0.00 \\ \hline
512/1024&0.99	&0.73	&0.00	              & 1.00   &  0.86  &  0.00         &0.99	&0.80	&0.00 \\ \hline
\end{tabular}
\end{center}
\caption{This table demonstrates
the errors and order accuracy of the softening parameter approach
for the $D_{2,2}$ disk for various number of zones $N=2^k$ from
$k=5$ to $10$.  
It shows that the order for the $D_{2,2}$ disk is about first order in $L^1$ and $L^2$ norm.
For measurement of $L^\infty$ norm, this method may fail in convergence under the pointwise sense.  
}
\label{tblSPCD2D2}
\end{table}

We implement the proposed method using MATLAB 7 software
under the computer system,  Intel Core 2 Duo CPU 1.8GHz with 2 GB RAM. 
The CPU time measurement information of the proposed method
is compared with the direct method in Table~\ref{tblCPUtime}. 
We list the CPU times in evaluating the kernels  ${\cal K}^{\cdot,\cdot}$, 
the force calculations of convolutions, and the whole process. 
The measurement is evaluated by the mean of 40 simulations.  It shows that the CPU times 
of both of the proposed method (P.M.) and the direct method (D.M.) are comparable. 
\begin{table}
\begin{center}
\begin{tabular}{|c|c|c||c|c||c|c|}  \hline
     & \multicolumn{2}{|c||}{Kernel ${\cal K}$}
     & \multicolumn{2}{|c||}{Force}
     & \multicolumn{2}{|c|}{The whole process}\\ \hline 
$N$  &P.M.      & D.M.    &  P.M.     & D.M.    & P.M.     & D.M.    \\ \hline 
  32 & 9.73E-3  & 7.43E-3 &  6.10E-3  & 3.43E-3 & 1.60E-2  & 2.31E-2 \\ \hline 
  64 & 3.80E-2  & 2.39E-2 &  2.08E-2  & 1.26E-2 & 5.87E-2  & 3.74E-2 \\ \hline 
 128 & 1.27E-1  & 9.67E-2 &  1.06E-1  & 6.43E-2 & 2.43E-1  & 1.60E-1 \\ \hline 
 256 & 5.11E-1  & 3.84E-1 &  6.48E-1  & 3.96E-2 & 1.18E+0  & 7.84E-1 \\ \hline 
 512 & 2.18E+0  & 1.57E+0 &  2.75E+0  & 1.61E+0 & 4.83E+0  & 3.29E+0 \\ \hline 
1024 & 8.59E+0  & 6.29E+0 &  1.13E+1  & 6.49E+0 & 2.01E+1  & 1.43E+1 \\ \hline 
\end{tabular}
\end{center}
\caption{This table demonstrates the CPU time measurement of
the proposed method (P.M.) and direct method (D.M.) with softening parameters.
The whole process consists of the generation of kernels
and the forces of calculations.  
It shows that the CPU times of both of P.M. and D.M. are comparable.}
\label{tblCPUtime}
\end{table}


\section{Discussion and conclusion}
\label{secConlusion}
We have presented a near second order method for calculating the self-gravitating force
of an infinitesimally thin disk for Cartesian coordinates.
For polar coordinates, we find that the method is near first order, $\sim 0.89$, only.
To quantify the accuracy, we define 
\begin{eqnarray*}
E_k = 
\left|
\int^{\theta_k}_{-\theta_k}\ln (1-\cos(\theta))d\theta-\frac{1}{2}(\ln(1-\cos(\theta_k))+\ln(1-\cos(-\theta_k))2(\theta_k), 
\right|
\end{eqnarray*}
where $\theta_k=1/2^k$. 
Table~\ref{tblSingular} reveals that the accuracy of the trapzoidal rule 
for the integration of the function $\ln (1-\cos(\theta))$ is nearly of first order.
With an improvement of the singular integration of
$\ln(1-\cos(\theta))$, the accuracy can be increased for the proposed method in polar coordinates.
\begin{table}
\begin{center}
\begin{tabular}{|c|c|c|c||c|c|c||c|c|c|}  \hline
 (Term, $k$ )         &  2      &  3      &  4     & 5    & 6     & 7      & 8    &  9   & 10   \\ \hline 
$E_k$	              & 2.86	& 1.73    &  1.07  & 0.55 & 0.34  & 0.20   & 0.11 & 0.06 & 0.03 \\ \hline
$\log_2(E_{k-1}/E_k)$ &         & 0.75    &  0.79  & 0.82 & 0.84  & 0.85   & 0.87 & 0.88 & 0.89 \\ \hline \hline
\end{tabular}
\end{center}
\caption{This table demonstrates the accuracy of the trapzoidal rule for 
the  integration of the function $\ln(1-cos(\theta))$ is 
near of first order $\sim 0.89$.
}
\label{tblSingular}
\end{table}

We note that the fast Fourier transform is only used to reduce the computational time.
For the practical computation, one can extend the range of the summation in (\ref{Fx0ji}). By setting 
$\sigma_{i',j'}=0$ whenever either $i'$ or $j'$ is in the range $-N+1$ to $0$, the value of any of 
the $F^{x,0}_{i,j}$ is unaffected. Furthermore, we can take $\sigma_{i',j'}$ to be
periodic since the sum (\ref{Fx0ji}) does not involve any values of $i'$ and $j'$ 
outside the first period.  We are also free to take ${\cal K}^{x,0}_{i-i',j-j'}$ periodic 
by defining it to be the periodic function that agrees with 
(\ref{K0iijj}) for $i-i'$ and $j-j'$ in the range $[-N+1,N]$
of the Green function. 

An important feature of our approach is that the boundary is not assumed to be periodic. 
Our approach is limited to the Cartesian and polar coordinates with uniform 
and logarithmic grid discretization, respectively, {which allows for rapid computation.} 
That is, the restriction of a convolution of two vectors provides the rapid computation, but it is 
restricted to a grid discretization that is either uniform or logarithmic. If the discretization is 
arbitrary, then the FFT is not suitable.

We point out that our method may be useful for gravity computations on a nested grid consisting of 
uniform grids having different grid spacing designed to resolve a central region with a finer grid. 
Such an approach would be complementary to the fast algorithm for solving the Poisson equation on 
a nested grid presented by Matsumoto and Hanawa \cite{Matsumoto03}. 

\section*{Acknowledgments}
This paper is dedicated to the memory of Professor C. Yuan, 
who initiated the project on the development of the Antares codes.
We thank the two anonymous referees for their valuable suggestions which significantly 
improved the presentation of our method in this paper.
The author C. C. Yen thanks the Institute of Astronomy and Astrophysics,
Academia Sinica, Taiwan for their constant support.


\section*{Appendix A: The calculation of the force in the $y$-direction in Cartesian coordinate}
Let 
\begin{eqnarray}
\label{eqn:K0A}
{\cal K}^{y,0}_{i-i',j-j'}
= \int\!\!\!\int_{D_{i',j'}} 
\frac{(\bar y-y_j)}{\left((\bar x-x_i)^2+(\bar y-y_j)^2\right)^{3/2}}d\bar xd\bar y,
\end{eqnarray}
\begin{eqnarray}
\label{eqn:KxA}
{\cal  K}^{y,x}_{i-i',j-j'}
 =  \int\!\!\!\int_{D_{i',j'}} 
\frac{(\bar y-y_j)(\bar x - x_{i'})}{\left((\bar x-x_i)^2+(\bar y-y_j)^2\right)^{3/2}}
d\bar xd\bar y,
\end{eqnarray}
and
\begin{eqnarray}
\label{eqn:KyA}
{\cal K}^{y,y}_{i-i',j-j'}
 =  \int\!\!\!\int_{D_{i',j'}} 
\frac{(\bar y-y_j)(\bar y - y_{j'})}{\left((\bar x-x_i)^2+(\bar y-y_j)^2\right)^{3/2}}
d\bar xd\bar y.
\end{eqnarray}
By (\ref{xforce}) and (\ref{sigmaapro}), we have 
\begin{eqnarray*}
{F}^y_{i,j} &\approx &\sum^N_{i'=1}\sum^N_{j'=1}
\int\!\!\!\int_{D_{i',j'}} 
\frac{\partial}{\partial y}{\cal K}(\bar x-x_i,\bar y-y_j,0)
\left(
\sigma_{i',j'} 
+\delta^x_{i',j'}(\bar x-x_{i'}) 
+\delta^y_{i',j'}(\bar y-y_{j'})
\right)d\bar x d\bar y\\
&:=& F^{y,0}_{i,j} + F^{y,x}_{i,j} + F^{y,y}_{i,j},
\end{eqnarray*}
where 
\begin{eqnarray}
\label{Fx0jiA}
F^{y,0}_{i,j} & = & \sum^N_{i'=1}\sum^N_{j'=1}\sigma_{i',j'}
\int\!\!\!\int_{D_{i',j'}} 
\frac{(\bar y-y_j)}{\left((\bar x-x_i)^2+(\bar y-y_j)^2\right)^{3/2}}d\bar xd\bar y
=\sum^N_{i'=1}\sum^N_{j'=1}\sigma_{i',j'} {\cal  K}^{y,0}_{i-i',j-j'},\\
\label{Fxx0jiA}
F^{y,x}_{i,j} & = & \sum^N_{i'=1}\sum^N_{j'=1}\delta^x_{i',j'}
\int\!\!\!\int_{D_{i',j'}} 
\frac{(\bar y-y_j)(\bar x - x_{i'})}{\left((\bar x-x_i)^2+(\bar y-y_j)^2\right)^{3/2}}d\bar xd\bar y
=\sum^N_{i'=1}\sum^N_{j'=1}\delta^x_{i',j'} {\cal  K}^{y,x}_{i-i',j-j'},\\
\label{Fxy0jiA}
F^{y,y}_{i,j} & = & \sum^N_{i'=1}\sum^N_{j'=1}\delta^y_{i',j'}
\int\!\!\!\int_{D_{i',j'}} 
\frac{(\bar y-y_j)(\bar y - y_{j'})}{\left((\bar x-x_i)^2+(\bar y-y_j)^2\right)^{3/2}}d\bar xd\bar y
=\sum^N_{i'=1}\sum^N_{j'=1}\delta^y_{i',j'} {\cal K}^{y,y}_{i-i',j-j'}.
\end{eqnarray}
The evaluation of (\ref{eqn:K0A}), (\ref{eqn:KxA}) and (\ref{eqn:KyA})
can be obtained with the help of the following simple integrals,
\begin{eqnarray*}
\int\!\!\!\int \frac{y}{(x^2+y^2)^{3/2}}dxdy  = -\ln (x+\sqrt{x^2+y^2})+C,\quad\quad\>
\int\!\!\!\int \frac{xy}{(x^2+y^2)^{3/2}}dxdy = -\sqrt{x^2+y^2}+C,
\end{eqnarray*}
\begin{eqnarray*}
\int\!\!\!\int \frac{y^2}{(x^2+y^2)^{3/2}}dxdy
= x\ln (y+\sqrt{x^2+y^2})+C, \quad
\int\!\!\!\int \frac{1}{(x^2+y^2)^{3/2}}dxdy
= -\frac{\sqrt{x^2+y^2}}{xy}+C.
\end{eqnarray*}
The value ${\cal K}^{y,0}_{i-i',j-j'}$ is equal to 
\begin{eqnarray}
\label{K0iijj}
{\cal K}^0_{i-i',j-j'}=
-\ln 
\left(
(\bar x - x_i) +\sqrt{(\bar x-x_i)^2 + (\bar y -y_j)^2}
\right)
\left|^{x_{i'+\frac{1}{2}}}_{x_{i'-\frac{1}{2}}}
\left|^{y_{j'+\frac{1}{2}}}_{y_{j'-\frac{1}{2}}} 
\right.\right.
\end{eqnarray}
The calculation of ${\cal K}^{y,x}_{i-i',j-j'}$ and ${\cal K}^{y,y}_{i-i',j-j'}$ 
are split into two parts
by the identity $(\bar y-y_j)(\bar y-y_{j'}) = (\bar y-y_j)^2 + (\bar y - y_j)(y_j - y_{j'})$,
and $(\bar y-y_j)(\bar x-x_{i'}) = (\bar y-y_j)(\bar x-x_i) 
+ (\bar y - y_j)(x_i - x_{i'})$, respectively.
It follows that
\begin{eqnarray*}
{\cal K}^{y,x}_{i-i',j-j'}
& = &(y_j-y_{j'}) {\cal  K}^{y,0}_{i-i',j-j'}+ \left(
(\bar  x-x_i)\ln (\bar y-y_j+\sqrt{(\bar y-y_j)^2 +(\bar x-x_i)^2)} 
\right)
\left|^{x_{i'+\frac{1}{2}}}_{x_{i'-\frac{1}{2}}}
\left|^{y_{j'+\frac{1}{2}}}_{y_{j'-\frac{1}{2}}}
\right.\right.,\\
{\cal  K}^{y,y}_{i-i',j-j'}
& = & (x_i-x_{i'}) {\cal K}^{y,0}_{i-i',j-j'}+\left(
-\sqrt{(\bar y-y_j)^2 + (\bar x-x_i)^2} \right)
\left|^{x_{i'+\frac{1}{2}}}_{x_{i'-\frac{1}{2}}}
\left|^{y_{j'+\frac{1}{2}}}_{y_{j'-\frac{1}{2}}} 
\right.\right..
\end{eqnarray*}

\section*{Appendix B: Calculations of convolution of two vectors by FFT}
It is known that the FFT of a vector $u_n$, $n=-N,\ldots,N-1$ can be rewritten as  
\begin{eqnarray*}
\hat u_k = \sum^{N-1}_{n=-N}u_n e^{-j2\pi k n/2N}, \mbox{ for } k=-N,\ldots,N-1,
\end{eqnarray*}
and its inverse FFT is given by 
\begin{eqnarray*}
u_n = \frac{1}{2N}\sum^{N-1}_{k=-N} \hat u_k e^{j2\pi kn/2N}, \mbox{ for } n=-N,\ldots,N-1.
\end{eqnarray*}
Let us consider two vectors $u_n$, $n=0,\ldots,N-1$ and 
$v_n$, $n=-N+1,\ldots,N-1$ and their inner product 
\begin{eqnarray*}
w_k = \sum^{N-1}_{n=0} u_n v_{k-n},\mbox{ for } k=0,\ldots,N-1.
\end{eqnarray*}
We set $w_k=0$, $k=-N,\ldots,0$ and 
\begin{eqnarray*}
\sum^{N-1}_{k=-N} w_k e^{-j2\pi m k /2N} 
&=&\sum^{N-1}_{k=-N}\sum^{N-1}_{n=-N} u_n v_{k-n}  e^{-j2\pi m k /2N}\\
&=&\sum^{N-1}_{n=-N}u_n e^{-j2\pi mn/2N}\sum^{N-1}_{k=-N} v_{k-n} e^{-j2\pi m (k-n) /2N}\\
&=&\sum^{N-1}_{n=-N} u_n e^{-j2\pi m n /2N} \sum^{N-1}_{k=-N} v_k e^{-j2\pi m k /2N} 
\end{eqnarray*}
This gives us 
\begin{eqnarray*}
\hat w_m = \hat u_m \cdot \hat v_m, \mbox{ for } m=-N,\ldots,N-1.
\end{eqnarray*}
Applying the inverse FFT on the above equation, we recover the vector $w_m$, $m=-N,\ldots,N-1$.
The vector $w_m$, $m=0,\ldots,N-1$ is the desired result.
\end{document}